\documentclass[paper]{JHEP3}

\usepackage{epsfig,multicol}

\newcommand\fverb{\setbox\pippobox=\hbox\bgroup\verb}
\newcommand\fverbdo{\egroup\medskip\noindent%
                        \fbox{\unhbox\pippobox}\ }
\newcommand\fverbit{\egroup\item[\fbox{\unhbox\pippobox}]}
\newbox\pippobox

\def\a{\alpha}
\def\b{\beta}
\def\c{\gamma}
\def\d{\delta}
\def\e{\epsilon}

\def\k{\kappa}
\def\l{\lambda}
\def\m{\mu}
\def\n{\nu}

\def\r{\rho}
\def\s{\sigma}
\def\t{\tau}

\def\w{\omega}

\def\C{\Gamma}
\def\D{\Delta}

\def\tr{{\rm tr}}

\def\Dslash{\,{\raise.15ex\hbox{/}\mkern-12mu D}}

\title{Strong Equivalence, Lorentz and CPT Violation, Anti-Hydrogen Spectroscopy
and Gamma-Ray Burst Polarimetry  \thanks{This research is supported in part by 
PPARC grant PP/G/O/2002/00470.  }}

\author{Graham M. Shore\\
        Department of Physics\\
        University of Wales, Swansea\\
        Swansea SA2 8PP, U.K.\\
        E-mail: \email{g.m.shore@swansea.ac.uk}}

\preprint{SWAT 04-410}
 
\abstract{The strong equivalence principle, local Lorentz invariance and CPT symmetry are
fundamental ingredients of the quantum field theories used to describe elementary particle
physics. Nevertheless, each may be violated by simple modifications to the dynamics
while apparently preserving the essential fundamental structure of quantum field theory
itself. In this paper, we analyse the construction of strong equivalence, Lorentz and
CPT violating Lagrangians for QED and review and propose some experimental tests in the 
fields of astrophysical polarimetry and precision atomic spectroscopy. In particular,
modifications of the Maxwell action predict a birefringent rotation of the direction of
linearly polarised radiation from synchrotron emission which may be studied using
radio galaxies or, potentially, gamma-ray bursts. In the Dirac sector, changes in atomic
energy levels are predicted which may be probed in precision spectroscopy of
hydrogen and anti-hydrogen atoms, notably in the Doppler-free, two-photon
$1s-2s$ and $2s-nd ~~(n \sim 10)$ transitions. 
}

\begin{document}

\section{Introduction}

The quantum field theories used to describe elementary particle physics incorporate
a number of basic principles including Lorentz invariance, unitarity, causality, locality 
and CPT symmetry and, in the presence of gravity, the weak and strong equivalence principles. 
Although there are no compelling theoretical or experimental reasons to question these
principles, it is nevertheless interesting to speculate on the experimental consequences
should any of them turn out to be violated in nature.

In this paper, we consider modifications of the dynamics of QED exhibiting either 
violations of the strong equivalence principle or of Lorentz and CPT invariance
and discuss certain experimental signatures, notably in the surprisingly
related fields of precision hydrogen and anti-hydrogen spectroscopy and polarimetry
of astrophysical sources such as radio galaxies or gamma-ray bursts.

The weak equivalence principle requires the existence at each spacetime point of a local
inertial frame of reference. This is fundamental to the structure of general relativity
and is realised by formulating spacetime as a (pseudo-)Riemannian manifold. We will keep
this as the basis for quantum field theory in curved spacetime. The strong equivalence
principle, however, apparently has a rather different status, being simply a restriction 
on the dynamics of the theory. It requires that the laws of physics (i.e.~dynamics)
are identical in each of these local inertial frames, where they reduce to their special
relativistic form. Precisely, this requires the matter fields to couple to gravity
through the connections only (minimal substitution), not through direct couplings to
the curvature tensors. Strong equivalence violation can therefore be realised by including 
in the QED Lagrangian interaction terms coupling the electromagnetic or Dirac fields directly 
to the curvature, e.g. ${1/M^2}R_{\m\n\l\r}F^{\m\n}F^{\l\r}$ or  ${1/M^2}R_{\m\n}\bar\psi 
\c^\m D^\n \psi$.  $M$ is a characteristic mass scale for strong equivalence
violation and is to be determined, or bounded, from experiment. 

In the electromagnetic sector, it is known from the original work of Drummond and 
Hathrell\cite{DH}
that terms of this type arise in the effective action for QED in curved spacetime as a 
result of integrating out electron loops. In this case, the mass scale $M$ is simply the 
electron mass $m$ and the coefficients are of order $\a$, the fine structure constant. 
In previous work\cite{DH,Sone,Stwo,Sthree,Sfour,Sfive}, 
we have investigated at length the implications of this effective 
quantum-induced violation of the strong equivalence principle for photon propagation. 
By far the most interesting result is the apparent prediction of superluminal propagation of
light. A careful review of this effect and its implications for causality has been given in 
refs.\cite{Sfive,Sseven}, 
where it is shown that in general relativity, superluminal propagation may in fact be 
compatible with stable causality. The situation is complicated by the fact that the
full effective action for QED in curved spacetime\cite{Sfive,Ssix}
involves interactions and 
form factors depending on spacetime covariant derivatives in addition to the terms in 
the original Drummond-Hathrell action, e.g. ${1/M^4}R_{\m\n\l\r}D_\s F^{\s\r} D^\l F^{\m\n}$.
This implies that light propagation is dispersive and the analysis of causality requires 
the identification of the relevant `speed of light', which turns out to be identifiable as 
the high-frequency limit of the phase velocity\cite{Leon}. 

In section 2, we construct a general strong equivalence violating QED Lagrangian in both
the photon and electron sectors, keeping interactions of all orders in derivatives but,
as above, only of first order in the gravitational curvature. The photon part of the
Lagrangian is essentially the same as the effective action derived in ref.\cite{Ssix}.
The electron Lagrangian, however, is quite new, although related to previous work on the
propagation of massless neutrinos in a background gravitational field\cite{Ohkuwa}.

It is important to realise that, unlike the photon case, this electron Lagrangian does
not have an interpretation as a low-energy effective action for QED -- there is no way
of `integrating out the photon field' to leave a local, infra-red finite effective action
describing electron propagation in curved spacetime. Nevertheless, we can relate our
strong equivalence violating generalisation of the Dirac Lagrangian to previous analyses
of the electron propagator in curved spacetime (more precisely, the evaluation of the
electron matrix elements of the energy-momentum tensor) incorporating one-loop 
self-energy contributions\cite{BG,Milton,DHGK}

Lorentz and CPT violating dynamics is introduced in a similar way. We begin as usual
with the standard formulation of QED in Minkowski spacetime, but include tensor
operators in the action multiplied by multi-index coupling constants, e.g.
$K_{\m\n\l\r}F^{\m\n}F^{\l\r}$ or $c_{\m\n}\bar\psi \c^\m D^\n \psi$. Since these
couplings are simply collections of constants, not tensor fields in their own right,
Lorentz invariance is explicitly violated in the dynamics of this modified QED.
Although it is not necessary for this phenomenological approach, we can imagine
these couplings to be Lorentz-violating VEVs of tensor fields in some more
fundamental theory. This approach to Lorentz and CPT violation has been pioneered
by Kosteleck\'y\cite{CK}, 
who has constructed the most general Lorentz-violating generalisation
of the standard model incorporating renormalisable (dimension $\le 4$) operators.
In an extended series of papers, Kosteleck\'y and collaborators have examined in 
detail a great variety of potential experimental signatures of these new
interactions. For a review, see, e.g.~ref.\cite{Kreview}.

The analogy between these Lorentz and CPT violating Lagrangians and the strong
equivalence violating QED Lagrangian is clear, with, for example, the coupling
constant $K_{\m\n\l\r}$ playing the role of the Riemann tensor $R_{\m\n\l\r}$.
It is therefore easy to translate aspects of the phenomenology of one theory
to the other. As we shall see in section 4, this leads to some new insights 
and economy of analysis.

A novel feature of the Kosteleck\'y Lagrangian is its inclusion, as a subset of the 
Lorentz-violating terms, of interactions that are CPT odd. We recall that CPT
invariance is a general theorem of quantum field theory assuming the general
principles of Lorentz invariance, locality and microcausality are respected (see, 
e.g.~ref.\cite{Weinberg}.
The Kosteleck\'y Lagrangian evades the CPT theorem by explicitly breaking 
Lorentz invariance. It therefore provides an interesting arena in which to
study the possible experimental signatures of CPT violation. This is particularly
important at present in view of the current experiments on anti-hydrogen
spectroscopy which are designed to test the limits of CPT with unprecedented
accuracy.

In the second part of this paper, we turn to some possible experimental consequences
of these novel phenomenological Lagrangians. As we have already extensively
investigated, the strong equivalence violating modifications to the QED
Lagrangian have important consequences for photon propagation. The light cone
is modified and in many cases superluminal propagation becomes possible.
Moreover, if the Weyl tensor $C_{\m\n\l\r}$ (the trace-free part of the Riemann
tensor) is non-vanishing, the effective light cone acquires a dependence
on the polarisation, so that the speed of light becomes polarisation dependent.
This is the phenomenon of gravitational birefringence. We immediately realise
that the same birefringent effect will also arise with the Lorentz-violating 
Lagrangian\cite{CK}, provided the coupling $K_{\m\n\l\r}$ has a trace-free component
analogous to $C_{\m\n\l\r}$. The Newman-Penrose formalism, which was found to
be an elegant way of analysing results in the strong equivalence violating
case, may similarly be usefully adapted to the Lorentz-violating Lagrangians.

One of the most important physical manifestations of birefringence is the rotation 
of the plane of linearly polarised light as it passes through the birefringent
medium, in our case either the background gravitational field or, in the
Lorentz-violating scenario, simply the vacuum. Since we expect any strong equivalence,
Lorentz or CPT violations to be tiny, detection of this birefringent rotation
requires experiments with the longest possible baseline, since the
rotation angle will be proportional to the distance the light has propagated through 
the birefringent medium. This leads us to focus on astrophysical sources which emit
radiation with a well-defined orientation of linear polarisation. A thorough
analysis of this effect for the case of the CPT-violating Chern-Simons interaction
$L^\m \e_{\m\n\l\r}A^\n F^{\l\r}$ in the Kosteleck\'y Lagrangian was performed
some time ago by Carroll, Field and Jackiw\cite{CFJ}, who were able to bound
the coupling $L^\m$ by placing limits on such a birefringent rotation from
observations of synchrotron emission from radio galaxies. In section 4, we briefly
review this and more recent work by Kosteleck\'y and Mewes\cite{KM}, 
then go on to discuss the possibility of improving these bounds
by considering higher redshift sources, in particular by exploiting the
linear polarisation observed in the afterglow of gamma-ray bursts to look
for a Lorentz-violating birefringent rotation. 

Curiously, while looking for strong equivalence, Lorentz and CPT violation in
the photon sector requires experiments on astronomical scales, testing the 
electron sector is the domain of precision atomic spectroscopy. Even more
intriguingly, since symmetry violation in the photon sector implies, through
radiative corrections, a corresponding violation in the electron sector (and 
vice-versa), experimental bounds on the Kosteleck\'y photon couplings $K_{\m\n\l\r}$
and $L^\m$ and the electron couplings $a_\m, b_\n, \ldots, h_{\m\n}$ are
in principle correlated. That is, bounds on Lorentz and CPT violation found
from astrophysics may also constrain the violation of these symmetries in the
experiments on anti-hydrogen spectroscopy. However, there are some caveats
here related to subtle anomaly-related issues in QED\cite{JK}, so the situation is
actually not quite so clear-cut (see section 3).

In section 5, therefore, we address the question of strong equivalence, Lorentz
and CPT violation in the electron sector of QED, using the appropriately
modified Dirac equation to recalculate atomic energy levels and their
implications for precision hydrogen and anti-hydrogen spectroscopy.
Following the production of significant numbers of cold anti-hydrogen
atoms by the ATHENA\cite{ATHENA,ATHENAtwo} and ATRAP\cite{ATRAPone,ATRAPtwo} 
experiments at CERN, the potential of performing
precision measurements of the anti-hydrogen as well as hydrogen spectrum
may be realised, and it is hoped that experiments will soon be 
underway to test CPT to high accuracy
by comparing the frequencies of the $1s-2s$ transition in hydrogen and
anti-hydrogen. This transition is favoured because of the extremely narrow line
width of the $2s$ state, since the only available decay is the Doppler-free,
two-photon $1s-2s$ transition. However, within the framework of the Kosteleck\'y
Lagrangian, CPT violation would only be observable in the hyperfine Zeeman
splittings of this transition for trapped anti-hydrogen\cite{BKR}, 
where the line widths
would be subject to Zeeman broadening. Here, we present a slightly extended
analysis of the energy level shifts predicted by the modified Dirac equation
and propose an alternative measurement which in practice may
be of comparable accuracy, namely the Doppler-free, two-photon $2s-nd ~~
(n \sim 10)$ transition for hydrogen and anti-hydrogen.

\section{Strong Equivalence Violating QED}

The weak equivalence principle requires the existence of a local inertial frame
at each spacetime point. This is realised by making spacetime a (pseudo-)Riemannian
manifold, which admits a flat tangent space at each point, related to spacetime
via the vierbein $e^\m{}_a$ (where $a$ is the tangent space index).  In quantum
field theory in curved spacetime, the fundamental fields are defined on this tangent
space.

The strong equivalence principle is invoked to constrain the dynamics. It requires
that the laws of physics are the same at the origin of the local inertial frame
at each spacetime point, where they reduce to their usual special relativistic
form. This is realised by requiring the coupling of the fields to gravity to involve
only the spacetime connection, not directly the curvature tensors. It is equivalent
to minimal substitution, where the general relativistic equations of motion are
found from the special relativistic ones by substituting the covariant derivatives
for ordinary derivatives. The same principle is of course already used to construct
the usual QED Lagrangian itself -- the electron field $\psi$ is coupled to 
electromagnetism via the gauge covariant derivative only, with the (non-renormalisable)
operators coupling $\bar\psi \psi$ bilinears directly to the field strength 
$F_{\m\n}$ being omitted.  The conventional QED action is therefore simply
\begin{equation}
\int d^4x \sqrt{-g}~\biggl[~-{1\over4}F_{\m\n}F^{\m\n} ~+~ 
\bar\psi\Bigl(i e^\m{}_a \c^a D_\m   - m \Bigr) \psi ~\biggr]
\label{eq:ba}
\end{equation}
where $D_\m \psi = \bigl(\partial_\m  - {i\over2}\w_{\m ab}\s^{ab} + ieA_\m\bigr) \psi$
is the covariant derivative, involving the spin connection 
$\w_{\m ab} = e_{\l a}\bigl(\partial_\m e^\l{}_b + \C^\l_{\m\n} e^\n{}_b \bigr)$.
Here, $\s^{ab} = {i\over4}[\c^a,\c^b]$, and in what follows we shall usually abbreviate
$\c^\m = e^\m{}_a \c^a$.

\subsection{Extended Maxwell action}

The strong equivalence principle is violated by the introduction of direct
couplings to the curvature tensors $R_{\m\n\l\r}, R_{\m\n}$ and $R$, e.g.
$1/M^2 R_{\m\n\l\r} F^{\m\n} F^{\l\r}$. This implies that the local
dynamics now distinguishes between spacetime points, since it obviously depends 
on the curvature at each point. Since these new interactions necessarily involve
non-renormalisable (dim $> 4$) operators, their inclusion introduces
a new scale $M$, which in our phenomenological approach characterises the
scale of strong equivalence violation. Its value is to be determined, or bounded,
by experiment.

We will only consider operators which are linear in the curvature. This is most
naturally interpreted as keeping only the lowest-order terms in an expansion
of a more general Lagrangian in $O(R/M^2)$, so we regard the resulting theory
as a valid approximation for gravitational fields which are weak on the scale $M$.
On the other hand, we will write down an action which involves all orders in
covariant derivatives acting on the fields. In this case, keeping only terms
of lowest order in $O(D^2/M^2)$ is a low-momentum (or low-frequency) approximation.
This will nevertheless be a useful first step in understanding the phenomenology
of strong equivalence violation.

A systematic analysis of all possible $O(R/M^2)$ operators shows that the most
general strong equivalence violating extension of the Maxwell sector of QED at
this order has the action
\begin{eqnarray}
&\nonumber\\
&\C~~ =~~ \int d^4 x \sqrt{-g} \biggl[~ -{1\over4}F_{\m\n}F^{\m\n}~+~{1\over M^2}
\Bigl(D_\m F^{\m\l}~ \overrightarrow{d_0}~ D_\n F^\n{}_{\l}~~~~~~~~~~~~~~~~~~~~ 
\nonumber\\
&~~~~~~~~~+~\overrightarrow{a_0}~ R F_{\m\n} F^{\m\n}~ 
+~\overrightarrow{b_0}~ R_{\m\n} F^{\m\l}F^\n{}_{\l}~
+~\overrightarrow{c_0}~ R_{\m\n\l\r}F^{\m\n}F^{\l\r} \Bigr) \nonumber\\
&~~+~{1\over M^4}\Bigl(\overrightarrow{a_1}~ R D_\m F^{\m\l} D_\n F^\n{}_{\l}~  
+~\overrightarrow{b_1}~ R_{\m\n} D_\l F^{\l\m}D_\r F^{\r\n} \nonumber\\
&~~~~~~~~~+~~\overrightarrow{b_2}~ R_{\m\n} D^\m F^{\l\r}D^\n F_{\l\r}~
+~\overrightarrow{b_3}~ R_{\m\n} D^\m D^\l F_{\l\r} F^{\r\n}  \nonumber\\
&~~~~~~~~~~~~~~~~~+~\overrightarrow{c_1}~ R_{\m\n\l\r} D_\s F^{\s\r}D^\l F^{\m\n} ~
\Bigr)~~+~{1\over M^6}~\overrightarrow{b_4}~ R_{\m\n} D^\m D_\l F^{\l\s}D^\n 
D_\r F^\r{}_\s   ~~\biggr] 
\label{eq:bb}
\end{eqnarray}
In this formula, the $\overrightarrow{a_n}$, $\overrightarrow{b_n}$, 
$\overrightarrow{c_n}$ are form factor functions of three operators, i.e.
\begin{equation}
\overrightarrow{a_n} \equiv a_n\Bigl({D_{(1)}^2\over M^2}, {D_{(2)}^2\over M^2}, 
{D_{(3)}^2\over M^2}\Bigr)
\label{eq:bc}
\end{equation}
where the first entry ($D_{(1)}^2$) acts on the first following term
(the curvature), etc. $\overrightarrow{d_0}$ is similarly defined as a single variable 
function.

To establish eq.(\ref{eq:bb}), extensive use has been made of the Bianchi identities
for the electromagnetic field strength and the curvature tensors, viz.
$D_{[\s} F_{\m\n]} = 0, ~ D_{[\s} R_{\m\n]\l\r} = 0, ~ D^\m \bigl( R_{\m\n} - {1\over2}
R g_{\m\n}\bigr) = 0$, the commutation relation $[D_\m,D_\n] \sim O(R)$ for
covariant derivatives, and repeated use of integration by parts to show
the equivalence in the action of different operators. As a simple illustration,
notice for example that
\begin{equation}
\int d^4x~R_{\m\n} D^\m D^\n F^{\l\r} F_{\l\r} ~~=~~
- \int d^4x~R_{\m\n} D^\m F^{\l\r} D^\n F_{\l\r} ~
+~{1\over4}\int d^4x D^2 R F^{\l\r} F_{\l\r}
\label{eq:bd}
\end{equation}
so the apparently independent operator on the l.h.s. is equivalent to a combination
of the $b_2$ and $b_0$ terms, taking account of the form factors.\footnote{
Notice that in the effective action in ref.\cite{Sfive,Ssix},  
$\int d^4x~R_{\m\n} D^\m D^\n F^{\l\r} F_{\l\r}$ and 
$\int d^4x~R_{\m\n} D^\m F^{\l\r} D^\n F_{\l\r}$ were included as independent terms.
The effective action quoted there can therefore be simplified. They do, however,
arise from the quite different operators $\tr R_{\m\n} D^\m D^\n \hat P \hat P$
and $\tr R_{\m\n} D^\m \hat{\cal R}^{\l\r} D^\n \hat{\cal R}_{\l\r}$ in the
Barvinsky {\it et al.} action for general background fields \cite{BGVZone,BGVZtwo}. }

In applications of eq.(\ref{eq:bb}), in particular to the analysis of birefringence
in section 4, it is often convenient to introduce the Weyl tensor $C_{\m\n\l\r}$.
Indeed, it is only the Weyl tensor contribution which gives rise to birefringence.
Explicitly, $C_{\m\n\l\r}$ is the trace-free part of the Riemann tensor:
\begin{eqnarray}
&\nonumber\\
C_{\m\n\l\r} ~=~ R_{\m\n\l\r} ~-~&{1\over2}\bigl(R_{\m\l}g_{\n\r} - R_{\m\r}g_{\n\l}
- R_{\n\l}g_{\m\r} + R_{\n\r}g_{\m\l} \bigr) \nonumber\\
&+~{1\over6} R \bigl(g_{\m\l} g_{\n\r} - g_{\m\r} g_{\n\l} \bigr)~~~~~~~~~~
\label{eq:be}
\end{eqnarray}
In contrast to the Ricci tensor $R_{\m\n}$, the Weyl tensor is not directly constrained
by matter via the Einstein field equations. The two operators involving the Riemann
tensor in eq.(\ref{eq:bb}) can be re-expressed in terms of the Weyl tensor
and combinations of the Ricci tensor and Ricci scalar terms as follows. (Note that 
these identities hold only under the integral and to $O(R)$.)
\begin{equation}
\int d^4x \sqrt{-g}~ C_{\m\n\l\r} F^{\m\n}F^{\l\r} = 
\int d^4x \sqrt{-g}~ \biggl[ R_{\m\n\l\r} F^{\m\n}F^{\l\r}
- 2 R_{\m\n} F^{\m\l} F^\n{}_\l  + {1\over3} R F^{\m\n} F_{\m\n}\biggr]
\label{eq:bf}
\end{equation}
and 
\begin{eqnarray}
&\nonumber\\
\int d^4x \sqrt{-g}~ C_{\m\n\l\r}D_\s F^{\s\r} D^\l F^{\m\n} = 
\int d^4x \sqrt{-g}~ \biggl[R_{\m\n\l\r}D_\s F^{\s\r} 
D^\l F^{\m\n} ~~~~~~~~~~~~~~~~~~~~~~~~~~~~ 
\nonumber\\
- R_{\m\n} D_\l F^{\l\m} D_\r F^{\r\n} 
- R_{\m\n} D^\m D^\l F_{\l\r} F^{\r\n}~~~ \nonumber\\
- {1\over6} R D_\m F^{\m\l} D_\n F^{\n\l} - 
{1\over4} R D^2 F^{\l\r} F_{\l\r}\biggr]~~~~~~~
\label{eq:bg}
\end{eqnarray}

While in this paper we are considering eq.(\ref{eq:bb}) as a phenomenological Lagrangian
exhibiting violation of the strong equivalence principle, it is important to realise
that the same expression (with $M$ interpreted as the electron mass) arises in 
conventional QED in curved spacetime as the effective action for the electromagnetic 
field when we include one-loop vacuum polarisation to integrate out the electron field. 
It is therefore the appropriate action to use to study photon propagation in curved 
spacetime, taking into account quantum effects at $O(\a)$. There is just one 
difference -- at one-loop, the quantum effective action does not have the $O(1/M^6)$ 
term that we have included above. Exact expressions for the form factors derived in 
QED at $O(\a)$ are known and are given in ref.\cite{Ssix}.

\subsection{Extended Dirac action}

A similar analysis can be carried out to find the most general free Dirac action
comprising strong equivalence breaking operators of first order in the curvature
but all orders in derivatives. A systematic study of all the possibilities shows
that the following is a complete basis:
\begin{eqnarray}
&\nonumber\\
&\C ~~=~~ \int d^4 x \sqrt{-g} ~\biggl[\bar\psi \bigl(i\c.D - m\bigr)\psi
+ {1\over M}~\bar\psi ~\overrightarrow h_1~ D^2 \psi  
+ {1\over M^2}~i\bar\psi ~
\overrightarrow h_2~ D^2 \c.D\psi ~~~~~~~~~~~~~~~~~~~ \nonumber\\
&+ {1\over M}~\overrightarrow f_1~ R \bar\psi \psi
+ {1\over M^2}~\Bigl(~\overrightarrow f_2~ iR \bar\psi \c.D\psi
+ \overrightarrow f_3~ iR D_\m\bigl(\bar\psi\c^\m \psi\bigr)~ \Bigr)
+ {1\over M^3} \overrightarrow f_4~ iR D_\m\bar\psi \s^{\m\n} D_\n\psi ~~~~~~~~~ 
\nonumber\\
&+ {1\over M^2}~ \overrightarrow g_1~ iR_{\m\n} \bar\psi \c^\m D^\n\psi
+ {1\over M^3}~\Bigl(~\overrightarrow g_2~ R_{\m\n}D^\m\bar\psi D^\n\psi
+ \overrightarrow g_3~ 
iR_{\m\n} D^\m\bar\psi \s^{\n\l} D_\l\psi~\Bigr) ~~~~~~~~~~~~~~~~~~~
\nonumber\\
&+ {1\over M^4}~ \Bigl(~\overrightarrow g_4~ iR_{\m\n} D^\m\bar\psi \c.D D^\n\psi 
+ \overrightarrow g_5~ iR_{\m\n} D_\l\bigl(D^\m\bar\psi \c^\l D^\n\psi\bigr)~
\Bigr) ~~~~~~~~~~~~~~~~~~~~~~~~~~~~~~~~~~~~~~~ \nonumber\\
&+ {1\over M^5}~ \overrightarrow g_6~ 
iR_{\m\n} D^\m D_\l\bar\psi \s^{\l\r} D^\n D_\r\psi  ~~+{\rm h.c.}~~
\biggr]~~~~~~~~~~~~~~~~~~~~~~~~~~~~~~~~~~~~~~~~~~~~~~~~~~~~~~~~~~~~~
\label{eq:bh}
\end{eqnarray}
where as before the $\overrightarrow f_n$,  $\overrightarrow g_n$ and  
$\overrightarrow h_n$ are form factor functions of derivatives.

A notable feature of eq.({\ref{eq:bh}) is that there is no independent term
involving the Riemann tensor. That is, there is no possible term in the action
that can be built from the trace-free Weyl tensor. This follows relatively
straightforwardly from the symmetries of $R_{\m\n\l\r}$ and the (Bianchi) identity
$D^\m R_{\m\n\l\r} = D_\l R_{\n\r} - D_\r R_{\n\l}$. The independence of this 
Dirac action on the Weyl tensor is in marked contrast to the Maxwell action.

At first sight, it might be thought that just as the extended Maxwell action can 
be realised as the quantum effective action governing photon propagation in the 
presence of vacuum polarisation, this extended Dirac action could similarly be 
realised as an effective action for electrons incorporating one-loop self-energy
corrections in the presence of gravity. In fact, there is no local effective action
which accomplishes this\cite{DHGK}. 
The reason is that whereas it makes sense to write a 
low-energy effective Lagrangian valid below the scale of the electron mass, it is
not possible to do the same with the massless photon. If we try, we find that the
the form factors $f_i, g_i, h_i$ in the Dirac action cannot be local, 
i.e.~polynomial functions in $D^2$. As we see below, their Fourier transforms
include infra-red singular logarithms of momentum.

On the other hand, a slight generalisation of eq.(\ref{eq:bh}) to chiral fermions
{\it is} a good effective action for the propagation of neutrinos in curved
spacetime. In this case, the relevant one-loop self-energy diagrams involve the W
or Z boson propagators and the quantum corrections can be encoded at low energies
by an effective action for the neutrinos alone, with the strong equivalence breaking
scale $M$ being identified with the weak scale $m_W$. An analysis of neutrino
propagation in curved spacetime has been carried out some time ago by 
Ohkuwa\cite{Ohkuwa}, generalising the results of Drummond and Hathrell\cite{DH} 
for photons.  To lowest order
in curvature and for low-momentum neutrinos, he finds that the $\overrightarrow h_1, 
\overrightarrow f_2, \overrightarrow f_3$ and $\overrightarrow g_1$ operators, 
modified to include left-handed chiral propagators and with constant
coefficients of $O(\a/m_W)$, are sufficient to encode the one-loop self-energy 
corrections, and computes the $g_1(0,0,0)$ coefficient (the only one which affects
the neutrino velocity). Just as in the photon case, this allows the possibility of
superluminal propagation of (massless) neutrinos in certain spacetimes, notably
the FRW universe. However, whereas there is a birefringent shift in the photon
velocity (with one polarisation being superluminal) for Ricci-flat spacetimes such 
as the Schwarzschild black hole, the neutrino velocity remains equal to the speed of
light in this case. Our construction of the general Dirac action (\ref{eq:bh}) 
shows that in fact this is a general result, valid beyond the low-momentum 
approximation.

Eq.(\ref{eq:bh}) can easily be extended to include electron-photon interactions by
promoting the spacetime covariant derivatives to be gauge covariant as well. As 
explained above, this is the conventional prescription of minimal substitution which, 
in ordinary QED, ensures a renormalisable Lagrangian. It could reasonably be argued, 
however, that in the spirit of constructing (non-renormalisable) strong equivalence
violating phenomenological extensions of QED, we should also allow violation of
the gauge minimal substitution principle as well and include operators coupling the
electron bilinears $\bar\psi \psi$ directly to the electromagnetic field strength
$F_{\m\n}$ as well as the spacetime curvatures.  Examples of such interactions
would be $F_{\m\n} i\bar\psi \c^\m D^\n \psi$ or $F_{\m\n} \bar\psi \s^{\m\n} \psi$.
Since we will not make use of such interactions in this paper, however, we will not
present a classification of all such possibilities here.

\subsection{Energy-momentum tensor}

After reviewing the analogous formalism for broken Lorentz and CPT violation in the
next section, we shall discuss the implications of these extended Maxwell and Dirac
actions for the propagation of polarised light and for atomic spectroscopy in sections
4 and 5 respectively.

First, though, we shall evaluate the electron matrix element of the energy-momentum
tensor using the Dirac action (\ref{eq:bh}). This is an important object in several 
contexts, ranging from the study of general relativity as a low-energy effective theory
of quantum gravity\cite{DHGK} 
to deep inelastic scattering, where the energy-momentum tensor arises
in the operator product expansion of electromagnetic currents and the form factors 
are identified as structure functions (the matrix elements being taken between proton
rather than electron states). It is also important in determining the explicit
expressions for the form factors in any application where the extended Dirac action
can be viewed as a conventional low-energy effective action, e.g.~in the generalisation
to neutrino propagation. So while the discussion that follows is rather formal, the
results may have applications in a number of interesting contexts.

The energy-momentum tensor matrix element in flat spacetime is defined as
\begin{equation}
\langle p'| T^{\m\n} |p\rangle ~~=~~\bar u(p')~ {(-2)\over \sqrt{-g}}~
{\d^3\C\over \d\bar\psi \d g_{\m\n} \d\psi}\bigg|_{{\rm F.T.},~g_{\m\n}=\eta_{\m\n}} u(p)
\label{eq:bi}
\end{equation}
which follows immediately from the definition of $T^{\m\n}$ as the functional derivative
of the action with respect to the metric. Interpreted in perturbative quantum gravity,
this is the electron-graviton vertex.  

Its evaluation therefore amounts to taking the metric variation of the various terms
in the extended Dirac action. We therefore collect here some useful formulae for 
functional derivatives:
\begin{eqnarray}
\nonumber\\
&{\d\over\d g_{\m\n}(x)}~\C^\l_{\r\s}(y) ~~=~~\biggl[-{1\over2} g^{\l\{\m}g^{\n\}}{}_\r
D_\s \d(x,y) -{1\over2} g^{\l\{\m}g^{\n\}}{}_\s D_\r \d(x,y) \nonumber\\
&~~~~~~~~~~~~~~~~~~~~~~ + {1\over2} g^{\l\t} g^{\{\m}{}_\r g^{\n\}}{}_\s D_\t \d(x,y) 
\biggr]
\label{eq:bj} \\
&{\d\over\d g_{\m\n}(x)}~ R(y) ~~=~~ - \Bigl( R^{\m\n} + g^{\m\n} D^2 - D^\m D^\n \Bigr)
\d(x,y)~~~~~~~~~~~~~~~~~~~~~~~~~~~ 
\label{eq:bk} \\
&{\d\over\d g_{\m\n}(x)}~ R_{\l\r}(y) ~~=~~ -{1\over2}\Bigl( g^\m{}_\l g^\n{}_\r D^2
- g^\n{}_\r D_\l D^\m - g^\m{}_\l D_\r D^\n + g^{\m\n} D_\r D_\l \Bigr) \d(x,y)~~~~ 
\label{eq:bl} 
\end{eqnarray}
where, importantly, the derivatives are w.r.t.~$x$.

Acting on spinor quantities, the metric derivative is to be re-interpreted in terms
of the vierbein as
\begin{equation}
{\d\over\d g_{\m\n}} ~~\rightarrow~~ {1\over4} \biggl(e^\m{}_c {\d\over\d e_{\n c}}
+ e^\n{}_c {\d\over\d e_{\m c}} \biggr)
\label{eq:bm}
\end{equation}
Recalling that the vanishing of the covariant derivative of the vierbein
\begin{equation}
D_\l e^\r{}_b ~\equiv~ \partial_\l e^\r{}_b + \C^\r{}_{\l\s} e^\s{}_b
- \omega_{\l bc}e^{\r c} ~=~0
\label{eq:bn}
\end{equation}
defines the spin connection as
\begin{equation}
\omega_{\m ab} ~=~e_{\l a} \Bigl(\partial_\m e^\l{}_b + \C^\l_{\m\n}e^\n{}_b\Bigr)
~\equiv~ e_{\l a} \tilde D_\m e^\l{}_b
\label{eq:bo}
\end{equation}
where $\tilde D$ is covariant only w.r.t.~the curved spacetime index, we can also 
show that
\begin{equation}
{1\over4} \biggl(e^\m{}_c(x) {\d\over\d e_{\n c}(x)} + e^\n{}_c(x) 
{\d\over\d e_{\m c}(x)} \biggr)~
\omega_{\l ab}(y) ~~=~~{1\over2} \tilde D_\l \Bigl(e^{\{\m}{}_a e^{\n\}}{}_b~ \d(x,y)
\Bigr)
\label{eq:bp}
\end{equation}
where, again, the derivative is w.r.t.~$x$. Note that in this formalism,
the connection $\C^\l_{\m\n}$ is considered as independent of the vierbein $e^\m{}_a$.
With this interpretation, we can then readily check that the energy-momentum tensor
for the conventional Dirac action is simply
\begin{eqnarray}
&\nonumber\\
&T^{\m\n} ~~=~~ {-2\over\sqrt{-g}} ~{\d\over\d g_{\m\n}}~\int d^4 x \sqrt{-g}~
\bar\psi \bigl(i\c.D - m\bigr)\psi ~+~{\rm h.c.} \nonumber\\
&=~~{i\over2} \bar\psi \c^{\{\m} \overrightarrow D^{\n\}} \psi 
- g^{\m\n} \bar\psi \bigl(i\c.\overrightarrow D - m\bigr)\psi ~~~~~~
\label{eq:bq} 
\end{eqnarray}

The matrix elements of each of the terms in eq.(\ref{eq:bh}) can now be taken in turn.
For those already involving a curvature tensor, the only contributions to the flat 
spacetime matrix elements are clearly those arising from the variations of the curvature
tensors themselves using eqs.(\ref{eq:bk}),(\ref{eq:bl}). For these terms, therefore,
we do not need to take the metric variations of the form factors themselves. We do, however,
need to take into account the variations of the $\overrightarrow h_i$ form factors. 

The matrix element $\langle p'| T^{\m\n}(q) |p\rangle$, or electron-graviton vertex, can be
easily shown to have only three possible independent Lorentz structures in momentum space.
Each of these individually satisfies the conservation constraint $q_\m T^{\m\n} = 0$,
which follows from diffeomorphism invariance of the action, using the equations of motion
for the on-shell wave functions $u(p)$, $\bar u(p')$. (The kinematical variables used here
are $p_\m~(p'_\m)$ for the initial (final) electron momenta and $q_\m = p'_\m - p_\m$ 
for the momentum transfer (graviton momentum)). We can choose them to be
$(\c^\m P^\n + \c^\n P^\m), ~P^\m P^\n$ and $(q^\m q^\n - q^2 g^{\m\n})$, where $P_\m =
{1\over2}(p'_\m + p_\m)$. 
The form factors $\overrightarrow f_i$, $\overrightarrow g_i$ reduce on-shell to single 
functions of the momentum transfer squared, i.e.
\begin{equation}
\overrightarrow f_i ~\rightarrow ~ f_i\bigl(-q^2, -{p'}^2, -p^2\bigr)\Big|_{\rm on-shell}~~
\equiv ~~\tilde f_i(q^2)
\label{eq:br}
\end{equation}
To simplify notation, we also let $h_i(-q^2) -q^2 h'_i(-q^2) ~\equiv~ \tilde h_i(q^2)$.

Notice that the set of Lorentz structures chosen here is not unique. Another frequently
used basis replaces the term $\bigl(\c^\m P^\n + \c^\n P^\m\bigr)$ by 
${i\over m}\bigl(\s^{\m\l}q_\l P^\n + \s^{\n\l}q_\l P^\m\bigr)$. The translation between 
the two can be made using the Gordon identity
\begin{equation}
{i\over m}\bar u(p')~ \s^{\m\l}q_\l~ u(p) ~~=~~ \bar u(p')~\Bigl( \c^\m - {1\over m}P^\m 
\Bigr)~ u(p) 
\label{eq:bs}
\end{equation}

In order to simplify the analysis, we now choose to identify the strong equivalence
breaking scale $M$ with the electron mass $m$ so that there is only one mass scale in the
problem. This could easily be relaxed of course for a specific application. For example, 
for neutrino propagation we would take $M= m_W$ and neglect the neutrino mass compared
to the weak scale.
Putting all this together, we then find after an extensive calculation:
\begin{eqnarray}
&\nonumber\\
&\langle p'| T^{\m\n}(q) |p\rangle ~~=~~ \bar u(p')~\biggl\{
{1\over2}\bigl(\c^\m P^\n + \c^\n P^\m\bigr)~ \biggl[1 - \tilde h_2 ~ 
-~{1\over2} {q^2\over m^2}~ \tilde g_1 ~+~{1\over4}{q^2\over m^2}~ \tilde g_3 
\biggr]~~~~~~~~~~~~~\nonumber\\
&~~~~~~+ {1\over m} P^\m P^\n ~ \biggl[-2(\tilde h_1 + \tilde h_2) ~-~
{q^2\over m^2}~ \bigl( \tilde g_2 + \tilde g_4 + {1\over2} \tilde g_3 \bigr) ~-~
{1\over4} {q^4\over m^4}~ \tilde g_6 \biggr] \nonumber\\
&+ {1\over m} \bigl(q^\m q^\n - q^2 g^{\m\n} \bigr) ~ \biggl[(\tilde h_1 + \tilde h_2) 
~+~4 \bigl(\tilde f_1    + \tilde f_2  - {1\over4}{q^2\over m^2}~ \tilde f_4  \bigr) ~
\nonumber\\  
&~~~~~~~~~~~~~~~~~~~~~~~~~~~~~~~~~~~~~~~~~~~-~ {1\over2} {q^2\over m^2}~ \bigl( 
\tilde g_2 + \tilde g_4 + {1\over2} \tilde g_3 \bigr) ~+~ {1\over8} {q^4\over m^4}~ 
\tilde g_6 \biggr]~ \biggr\}~u(p) \nonumber\\ 
\label{eq:bt}
\end{eqnarray}
This expression shows clearly how the various possible terms in the general extended
Dirac action contribute to the three independent form factors in the energy-momentum
tensor matrix element.

Expressions of this type have appeared several times in the literature as  
the $O(h)$ contributions of one-loop electron self-energy diagrams in QED in 
the presence of a weak perturbation $g_{\m\n} = \eta_{\m\n} + h_{\m\n}$ of the Minkowski
spacetime metric, i.e.~as the QED corrections to the electron-graviton vertex in
perturbative quantum gravity. Explicit expressions for the functions $F_1(q^2), F_2(q^2)$ 
and $F_3(q^2)$ defined as the functions in square brackets in eq.(\ref{eq:bt}) have been 
given for QED by several authors\cite{BG,Milton,DHGK}. 
For small momentum transfer, the results are 
\begin{eqnarray}
&F_1 ~~=~~ 1 - {\a\over4\pi} {q^2\over m^2} \biggl(-{47\over18} + {\pi^2\over2}
{m\over\sqrt{-q^2}} + {2\over3} \ln {-q^2\over m^2} \biggr) ~+~\ldots 
\label{eq:bu}\\
&F_2 ~~=~~ {\a\over4\pi} {q^2\over m^2} \biggl( -{4\over9} -{\pi^2\over4}
{m\over\sqrt{-q^2}} - {4\over3} \ln {-q^2\over m^2} \biggr) ~+~\ldots ~~~~~~
\label{eq:bv}\\
&F_3 ~~=~~ {\a\over4\pi} \biggl( {11\over9} -{\pi^2\over2} {m\over\sqrt{-q^2}}
-{4\over3} \ln{-q^2\over m^2} \biggr) ~+~\ldots ~~~~~~~~~~
\label{eq:bw}
\end{eqnarray}
The $O(1)$ term in $F_1$, which follows directly from the free Dirac action, can readily
be seen to be a consequence of momentum conservation.

These expressions illustrate the essential problem in regarding the extended Dirac action
too literally as a low-energy effective action in the usual sense. The non-analytic
terms $\sqrt{-q^2}$ and $\ln(-q^2)$ are the signature of the long-range interactions in
QED mediated by the massless photon. These interactions clearly cannot be `integrated out'.
However, they do carry important physical information and, as shown recently in 
ref.\cite{DHGK}, can be used to reconstruct both classical and quantum corrections to
the Kerr-Newman and Reissner-Nordstr\"om metrics. 

The presence of non-analytic terms in the form factors for the energy-momentum tensor
matrix elements prevents us from using eqs.(\ref{eq:bu}),(\ref{eq:bv}) and (\ref{eq:bw})
to reconstruct polynomial form factors $\overrightarrow f_i$,  $\overrightarrow g_i$ and
$\overrightarrow h_i$ in the extended Dirac action and view it as a true local
effective action for QED in the same way as we have done for the extended Maxwell action. 
This programme could be carried out, however, with the chiral extension of the action where 
the scale $M$ is taken as the vector boson mass $m_W$ and the result interpreted as the
effective action for neutrino propagation. In this case, the self-energy quantum corrections
can indeed be encoded in an effective action of this type and a generalisation of
eq.(\ref{eq:bt}) could be used, as in the work of Ohkuwa\cite{Ohkuwa}, to constrain the
$\overrightarrow f_i$,  $\overrightarrow g_i$ and $\overrightarrow h_i$ form factors
by comparing with the results of explicit Feynman diagram calculations of the 
electron-graviton vertex in a Minkowski spacetime background.

\section{Lorentz and CPT Violating QED}

Lorentz invariance and CPT symmetry are fundamental properties of conventional
quantum field theories. Nevertheless, it is interesting to speculate that they may
not be exact in nature and to study possible signatures for their violation, if only
to stimulate increasingly precise experimental tests. A formalism for studying 
Lorentz and CPT violation within the framework of the standard model has been proposed
and extensively explored in recent years by Kosteleck\'y and collaborators\cite{CK,Kreview}. 
In this section, we briefly review this approach, emphasising the close technical 
similarities to our own work on strong equivalence violation.

In the Kosteleck\'y approach, explicit Lorentz violation is introduced by writing a
phenomenological Lagrangian comprising tensor operators with coupling constants carrying
spacetime indices. These couplings are simply collections of numbers, not tensor
fields in their own right. However, they may, though this is not at all necessary in the
phenomenological approach, be thought of as Lorentz-violating VEVs of tensor fields in
some more fundamental theory exhibiting spontaneous Lorentz violation. Restricting
to renormalisable interactions, the most general such extension of QED is therefore
(changing the notation of ref.\cite{CK} slightly):
\begin{eqnarray}
&\nonumber\\
&\C ~~=~~ \int d^4 x~~\biggl[- {1\over4} F_{\m\n} F^{\m\n} ~~+~~ \bar\psi\bigl(i \c.D
- m\bigr)\psi ~~~~~~~~~~~~~~~~~~~~~~~~~~~~~~~~~~~~~~~~~~~~~~~~~~\nonumber\\
&+ K_{\m\n\l\r} F^{\m\n} F^{\l\r}  ~-~ {1\over4} L^\m \e_{\m\n\l\r} A^\n F^{\l\r} 
~~~~~~~~~~~~~~~\nonumber\\
&+ ~\bar\psi\Bigl( - a_\m \c^\m ~-~ b_\m \c^5 \c^\m ~+~ i c_{\m\n} \c^\m D^\n ~+~
i d_{\m\n} \c^5 \c^\m D^\n ~-~ {1\over2} h_{\m\n} \s^{\m\n} \Bigr) \psi   \biggr]
\label{eq:ca}
\end{eqnarray}
With this notation, the couplings $a_\m$, $b_\m, L^\m$ of the super-renormalisable operators
have dimensions of mass while $c_{\m\n}, d_{\m\n}, h_{\m\n}, K_{\m\n\l\r}$ are
dimensionless. 

Here, we follow Kosteleck\'y and restrict attention to the simple renormalisable
Lagrangian (\ref{eq:ca}). Of course, nothing prevents us from introducing further
higher dimensional operators, particularly those with extra derivatives analogous to
eqs.(\ref{eq:bb}) and (\ref{eq:bh}). This would be in the spirit of regarding this
phenomenological action as a low-energy effective action in a theory in which
Lorentz symmetry is broken at a high scale $M$. 

Since Lorentz invariance is explicitly broken in (\ref{eq:ca}), one of the axioms of
the CPT theorem is not respected and therefore CPT symmetry is no longer guaranteed.
In fact, the super-renormalisable operators with couplings $a_\m, b_\n, L^\m$ are
CPT odd, while $c_{\m\n}, d_{\m\n}, h_{\m\n}$, $K_{\m\n\l\r}$ multiply CPT even
renormalisable operators. A phenomenological model of CPT violation is therefore obtained 
by using the Lagrangian (\ref{eq:ca}) with the couplings $a_\m, b_\n$ or $L^\m$
non-zero. It should also be noticed that in comparing results derived with positrons
and electrons, the substitutions $a_\m \rightarrow - a_\m$, 
$d_{\m\n} \rightarrow - d_{\m\n}$ and $h_{\m\n} \rightarrow - h_{\m\n}$ should be made
with the other couplings unchanged. 

Now, it is evident that the Lorentz-violating QED Lagrangian (\ref{eq:ca}) and the
strong equivalence violating extended QED Lagrangians (\ref{eq:bb}), (\ref{eq:bh}) have
many formal similarities. Tensor operators appear in each, though the role of the
curvature tensors in eqs.(\ref{eq:bb}), (\ref{eq:bh}) is played by the multi-index,
but non-tensorial, coupling constants in eq.(\ref{eq:ca}). Notice also that the Lagrangian
(\ref{eq:ca}) allows for parity-violating operators which we chose to exclude from
(\ref{eq:bb}), (\ref{eq:bh}). These would not occur in the quantum effective action
obtained from QED itself since this theory is parity-preserving, but
could be included in a purely phenomenological theory. It follows that the analysis of the 
phenomenological consequences of the two theories will be very similar, and many results
and predictions can simply be transcribed between them. We may also hope that some extra
insight may be gained from this comparison. The couplings which admit a direct equivalence
are:
\begin{eqnarray}
\nonumber\\ 
a_\m ~~ \sim ~~ f_3 D_\m R/M^2  \nonumber\\
c_{\m\n} ~~ \sim ~~ g_1 R_{\m\n}/M^2  \nonumber\\
K_{\m\n\l\r} ~~ \sim ~~ c_0 R_{\m\n\l\r}/M^2 
\label{eq:cb}
\end{eqnarray}
Of the others, $b_\m, d_{\m\n}$ and $L^\m$ are associated with the parity-violating 
operators which are not included in eqs.(\ref{eq:bb}), (\ref{eq:bh}). There is also no 
immediate analogue of the $h_{\m\n}$ term since $R_{\m\n}$ is symmetric and cannot 
couple to the operator $\bar\psi \s^{\m\n} \psi$ without higher derivatives.
 
Probably the most useful correspondence is of the set of couplings $K_{\m\n\l\r}$ with
the Riemann tensor $R_{\m\n\l\r}$. $K_{\m\n\l\r}$ inherits the algebraic symmetries of
the operator $F^{\m\n} F^{\l\r}$, which exactly matches the Riemann tensor. Just as we 
found it convenient to separate the Riemann tensor into the Ricci tensor and the
trace-free Weyl tensor, so it will be useful to split $K_{\m\n\l\r}$ in the same way,
defining the traced components $K_{\m\l}$ and $K$ and the trace-free components
$C_{\m\n\l\r}^{(K)}$ using the same formula (\ref{eq:be}) that defines the Weyl tensor. 
We will see in the next section, where we go further and introduce the Newman-Penrose 
formalism, that this is a useful analogy to exploit in the analysis of the birefringent 
propagation of light.

In the remainder of the paper, we investigate some of the experimental consequences of
this Lorentz and CPT violating Lagrangian, emphasising wherever possible the similarities,
and differences, with the strong equivalence violating curved spacetime Lagrangian.
First, though, we make some remarks about radiative corrections. It will be clear that
if we take the modified electron propagator from the Dirac sector of (\ref{eq:ca}) and
substitute it into the one-loop photon vacuum polarisation diagram, it will produce a
change in the photon propagator which can be realised as an $O(\a)$ contribution to
the coefficients of the Maxwell sector operators. And vice-versa, modifications of the
photon propagator induce $O(\a)$ contributions to the couplings of the Dirac sector 
operators. The result is that the electron and photon sector couplings are correlated
by $O(\a)$ radiative corrections. Exactly the same is of course true for the strong
equivalence violating action.

It follows that if we are able to put a bound on the Dirac couplings $a_\m, ~\ldots~, 
h_{\m\n}$ from experiments on atomic spectroscopy, then this will imply bounds on
the Maxwell couplings $K_{\m\n\l\r}, L^\m$ which are tested in astrophysical polarimetry 
experiments. And conversely, known bounds from astrophysics could imply bounds on future
precision hydrogen and anti-hydrogen spectroscopy experiments, such as those planned 
by ATHENA and ATRAP at CERN. The possibility that astrophysics and atomic 
spectroscopy experiments could be correlated in this way was one of the principal
motivations for this paper.

Perhaps the most interesting of these correlations would be those involving operators
which give rise to CPT violation in the anti-hydrogen spectrum. In particular, consider
the parity-violating Dirac coupling $b_\m$. At first sight, this appears to be linked
via radiative corrections with the coupling $L^\m$ of the Chern-Simons operator. Since,
as we see in the next section, the Chern-Simons term predicts a birefringent rotation
of linearly polarised light which could in principle be observed in the synchroton
radiation from radio galaxies, we are able to place bounds on $L^\m$ from astrophysics.
We would therefore expect this to imply a bound on $b_\m$, which would limit the size
of CPT violation which would be observed in the spectroscopic measurements at ATHENA
and ATRAP. 

Curiously though, the radiative correspondence between the axial-vector operator 
coupling $b_\m$ and the Chern-Simons coupling $L^\m$ is far from straightforward
theoretically. The question of whether a Chern-Simons term is induced via the 
one-loop vacuum polarisation diagram where the fermion propagator is modified to include
an axial-vector coupling turns out to be a subtle one in quantum field theory, involving
ambiguities associated with the axial anomaly. A very careful analysis of this issue has
been carried out by Jackiw and Kosteleck\'y\cite{JK}, who conclude that the 
question ``is a 
Chern-Simons term induced by vacuum polarisation in the theory with the Dirac action
modified by an axial-vector term'' has in fact no unique answer. Depending on how
we attempt to give a definition of the theory taking proper account of the axial anomaly,
a variety of reasonable answers may be given, including that the induced $L_\m = 
{3\a\over4\pi} b_\m$,~ $0$,~ or simply indeterminate.

In view of this interesting but indeterminate theoretical situation, it would therefore
be unwise to place too much confidence on a prediction of a bound for CPT violation in 
anti-hydrogen based on the bound from radio galaxies.

\section{Astrophysical Polarimetry}

We now turn to the phenomenological consequences of the strong equivalence, Lorentz
and CPT violating Lagrangians introduced so far. In this section, we focus on the
photon sector and in particular on the polarisation dependence of the propagation
of light (including radio waves and gamma-rays) in astrophysics.

\subsection{Light propagation and birefringence}

We begin with a brief review of light propagation based on the strong equivalence
violating Lagrangian (\ref{eq:bb}). This has been studied extensively in a series of
papers \cite{DH,Sone,Stwo,Sthree,Sfour,Sfive,Sseven} 
where we have explored the possibility of superluminal propagation and
its implications for causality as well as gravitational birefringence and dispersion.
The dispersive nature of light propagation, which as explained in ref.\cite{Sfive,Sseven} 
is essential in a precise analysis of causality and possible superluminal signal
propagation, only becomes apparent when the full effective action (\ref{eq:bb}) 
including the higher derivative operators is used. On the other hand, the basic features
of gravitational birefringence are already apparent in the low-momentum approximation 
to the action first derived by Drummond and Hathrell\cite{DH}. 
In this section, we will restrict 
ourselves to this simple case. This is also most directly analogous to the Kosteleck\'y
Lagrangian (\ref{eq:ca}) with its restriction to renormalisable operators only.

To study the propagation of light in this theory, we use the formalism of geometric
optics. The electromagnetic field is written in the form 
$A_\m = {\cal A} a_\m \exp{i\theta}$, where ${\cal A}$ is the amplitude and $a_\m$ is the
polarisation vector. The amplitude is taken to be slowly-varying on the scale of the
rapidly-varying phase $\theta$. The wave-vector (equivalent up to some 
subtleties\cite{Sseven} to the photon momentum) is identified as $k_\m = \partial_\m \theta$. 
The essential features of light propagation are then obtained from the equations of motion 
of the extended Maxwell action by considering the highest order terms in a controlled 
expansion in the rapidly-varying quantities. All this formalism
is explained in our previous papers (see \cite{Seight} for a pedagogical review).
The result is the following equation for the wave vector $k_\m$ and polarisation $a_\m$:
\begin{equation}
k^2 a^\n - k.a~ k^\n - {2b+4c\over M^2}~ R_{\l\r}\Bigl(k^\l k^\r a^\n - k^\l k^\n a^\r \Bigr)
- {8c\over M^2}~ C_{\m\l}{}^{\n}{}_{\r}k^\l k^\r a^\m ~~=~~ 0
\label{eq:da}
\end{equation}
where $a,b,c$ are coupling constants given from eq.(\ref{eq:bb}) as $a = a_0(0,0,0)$, etc.,
and we have explicitly introduced the Weyl tensor. The modified light cone follows
immediately from the condition
\begin{equation}
\det\biggl[k^2 g_{\m\n} - k_\m k_\n - {2b+4c\over M^2}\Bigl(R_{\l\r}k^\l k^\r g_{\m\n}
- R_{\l\m}  k^\l k_\n \Bigr) - {8c\over M^2}~C_{\m\l\n\r}k^\l k^\r \biggr] ~~=~~ 0
\label{eq:db}
\end{equation}

The physical light cone for photon propagation therefore no longer coincides with the
geometric light cones of the background curved spacetime, and so eq.(\ref{eq:db}) predicts 
that the speed of light will be different from the fundamental `speed of light' constant 
$c=1$. The remarkable feature implied by (\ref{eq:db}) is that in some cases it predicts 
that the speed of light may be greater than 1, i.e.~light propagation may be superluminal. 
For a careful discussion of what this means physically, we refer to our earlier 
papers\cite{Sfive,Sseven}. 
The second main prediction of eqs.(\ref{eq:da}), (\ref{eq:db}) is gravitational
birefringence, viz.~that the physical light cone (speed of light) will depend on the
polarisation if, and only if, the Weyl tensor is non-vanishing. See below for an explicit
demonstration.

As we have seen in earlier work, it is particularly useful in studying the implications
of eqs.(\ref{eq:da}), (\ref{eq:db}) to adopt the Newman-Penrose formalism. 
(See ref.\cite{Sthree}
for a relevant summary.) This involves introducing a null tetrad at each spacetime point
via a set of null vectors $\ell^\m, n^\m, m^\m, \bar m^\m$ satisfying 
$\l.n = 1,~ m.\bar m = -1,~ \ell.m = \ell.\bar m = n.m = n.\bar m = 0$. All tensors are
then specified by their components in this basis. In the discussion below, we will take
$\ell^\m$ as the direction of propagation, so that $k^\m = \omega \ell^\m$. 
$m^\m = {1\over\sqrt 2}\bigl(\tilde a^\m + i \tilde b^\m\bigr)$ is a complex linear 
combination of two spacelike vectors transverse to $\ell^\m$. It follows that $m^\m$ 
and $\bar m^\m$ represent respectively left and right-handed circular polarisation vectors, 
i.e.~the photon helicity $\pm 1$ eigenstates. The ten independent components of the Weyl 
tensor are described in the Newman-Penrose formalism by five complex scalars 
$\Psi_i$~ ($i = 0,\ldots,4$), where
\begin{eqnarray}
&\nonumber\\
&\Psi_0 = - C_{\m\n\l\r} \ell^\m m^\n \ell^\l m^\r ~~ \nonumber\\
&\Psi_1 = - C_{\m\n\l\r} \ell^\m n^\n \ell^\l m^\r ~~ \nonumber\\
&\Psi_2 = - C_{\m\n\l\r} \ell^\m m^\n \bar m^\l n^\r ~  \nonumber\\
&\Psi_3 = - C_{\m\n\l\r} \ell^\m n^\n \bar m^\l n^\r ~~ \nonumber\\
&\Psi_4 = - C_{\m\n\l\r} n^\m \bar m^\n n^\l \bar m^\r
\label{eq:dc}
\end{eqnarray}
with similar definitions for the ten independent components of the Ricci tensor.
Here, we only need the notation $\Phi_{00} = -{1\over2} R_{\m\n} \ell^\m \ell^\n$.

Applying this formalism to eq.(\ref{eq:da}), and choosing the polarisation to be
a linear combination $a^\m = \a m^\m + \b \bar m^\m$ of the left and right circular
polarisations, we find the matrix equation
\begin{equation}
\left(\matrix{\tilde k^2 & -{8c\over M^2}~\omega^2~ \Psi_0^* \cr
-{8c\over M^2}~\omega^2~ \Psi_0 &\tilde k^2 \cr} \right)~
\left(\matrix{\a \cr \b \cr} \right) ~~=~~ \left(\matrix{0 \cr 0 \cr} \right)
\label{eq:dd}
\end{equation}
where $\tilde k^2 = k^2 +{4b+8c\over M^2}~\omega^2 \Phi_{00}$ ~and we have used the
identity $C_{\m\n\l\r} \ell^\m m^\n \ell^\l \bar m^\r  = 0$. The eigenvalues
give the modified light cone\footnote{Eq.(\ref{eq:de}) corrects the corresponding
equation in ref.\cite{Sthree}, where a similar expression was quoted that, although
written in terms of $\Psi_0$ and $\Psi_0^*$, was only valid, and was only used,
for the case of real $\Psi_0$.}
\begin{equation}
k^2 = -{4b+8c\over M^2}~\omega^2~\Phi_{00} ~\pm~ {8c\over M^2}~\omega^2~|\Psi_0|
\label{eq:de}
\end{equation}
This implies a (non-dispersive) polarisation-dependent shift in the speed of light of
\begin{equation}
\d v ~~=~~ -{2b+4c\over M^2}~\Phi_{00} ~\pm~ {4c\over M^2}~|\Psi_0|
\label{eq:df}
\end{equation}
for the two polarisation eigenstates. We can now show that if we define a phase 
$\vartheta$ from the (complex) N-P scalar, $\Psi_{0} = |\Psi_0| e^{i\vartheta}$, 
then the eigenstates are 
\begin{equation}
a^\m ~~=~~ {1\over\sqrt 2}~\Bigl(e^{i{\vartheta\over2}}~ m^\m ~\pm 
e^{-i{\vartheta\over2}}~ \bar m^\m \Bigr)
\label{eq:dg}
\end{equation}
The states which propagate with a well-defined velocity are therefore superpositions
of the left and right circular polarisations with equal and opposite phases
determined by the Weyl tensor. They are therefore orthogonal linear polarisations,
the direction being determined by the Weyl phase.

It should now be reasonably clear that we can equally well apply this formalism to
the case of the Lorentz violating phenomenological theory in flat spacetime. 
(The formal discussion in the remainder of this section and in section 4.2 should
be compared with the treatment of photon propagation in refs.\cite{CK,KM}, where
similar issues are addressed without using the Newman-Penrose formalism. The main 
physics conclusions are the same.) 
As already noted, the coupling $K_{\m\n\l\r}$ appearing in the Kosteleck\'y Lagrangian
(\ref{eq:ca}) plays essentially the same role as the Riemann tensor in the above
discussion. In particular, it has the same number of independent components as
$R_{\m\n\l\r}$ by virtue of having the same algebraic symmetries, which are inherited
from the operator it multiplies in the Lagrangian. We can therefore introduce the
Newman-Penrose formalism in this context also and define N-P scalars $\Psi_i^{(K)}$~
$(i = 0,\ldots,4)$ from the trace-free analogue $C_{\m\n\l\r}^{(K)}$ of the Weyl
tensor (and similarly for the N-P analogues $\Phi_{ij}^{(K)}$ from
the traced components $K_{\m\n}$). This helps to put some order into the plethora
of components of $K_{\m\n\l\r}$ and extract the essential physics. All the
results derived above can now be directly translated. In particular, provided the
relevant components of $C_{\m\n\l\r}^{(K)}$ are non-vanishing, there will be a 
birefringent velocity shift analogous to eq.(\ref{eq:df}), viz.
\begin{equation}
\d v ~~=~~ -4~\Phi_{00}^{(K)} ~\pm~ 4~|\Psi_0^{(K)}|
\label{eq:dh}
\end{equation}
The velocity eigenstates are given by superpositions of the left and right
circular polarisations as in eq.(\ref{eq:dg}), where now $\vartheta$ is the phase
of $\Psi_0^{(K)}$. They are therefore orthogonal linear polarisations with orientation 
determined by $\vartheta$. 

Light with a general linear polarisation is a superposition of these eigenstates. 
As they propagate through some distance $D$, the eigenstates will pick up a phase 
difference $\d\phi = \pm {2\pi D\over \l} \d v$, where $\l$ is the wavelength. The 
resulting superposition is then no longer a pure linear polarisation, but will be
elliptically polarised.\footnote{This is the standard situation in optics of
superposing two orthogonal plane waves with a phase difference. The resulting figure 
swept out by the polarisation vector is in general an ellipse, but degenerates to a 
straight line for special values ($0, \pi, 2\pi, \ldots$) of the phase difference.
If the initial two waves have different amplitudes, then the orientation of the 
principal axes of the ellipse, as well as the ratio of its major to minor axes, 
depends both on this amplitude ratio as well as on the phase difference. The resulting
ellipses are the simplest examples of Lissajous figures. See, for example, 
ref.\cite{JW} for a detailed description.} For a generic
orientation of linear polarisation, the major axis of the ellipse will be rotated
relative to the direction of linear polarisation by an angle of $O(\d\phi)$. The ratio
of minor to major axes of the ellipse is also $O(\d\phi)$, the exact expressions in
each case depending in a non-trivial way on the angle between the polarisation
directions of the light wave and the velocity eigenstates. 
The transformation of linear to elliptic polarisation with a dependence on the
propagation distance $D$ is therefore a clear signal for Lorentz breaking through
a `Weyl' mechanism arising from the couplings $K_{\m\n\l\r}$ in the phenomenological
Lagrangian.

A similar analysis also applies to the Chern-Simons interaction in the Lagrangian
(\ref{eq:ca}). In this case, the appropriate equation of motion is
\begin{equation}
k^2 a^\n - k.a~ k^\n +i \e_{\m\n\l\r} L^\m k^\l a^\r ~~=~~ 0
\label{eq:di}
\end{equation}
The modified light cone is derived from
\begin{equation}
\det\biggl[k^2 g_{\m\n} - k_\m k_\n - i \e_{\m\n\l\r} L^\l k^\r \biggr] ~~=~~0
\label{eq:dj}
\end{equation}
which, for physical polarisations, implies\cite{CFJ}
\begin{equation}
k^4 + k^2 L^2 - (L.k)^2 ~~=~~0 
\label{eq:dk}
\end{equation}
To first order in $L^\m$, this gives a birefringent phase velocity shift\footnote{
Notice that, unlike the `Weyl' case, eq.(\ref{eq:dk}) implies a non-trivial
dispersion relation. In particular, while the phase velocity satisfies 
eq.(\ref{eq:dl}), the shift in the group velocity is second order in 
$L^\m$, i.e. ${\partial \omega\over\partial k} = 1 + O(L^2)$ \cite{CFJ}.} 
\begin{equation}
\d v ~~=~~ \pm {1\over2}~{1\over \omega^2} ~L.k
\label{eq:dl}
\end{equation}
The eigenstates themselves follow from eq.(\ref{eq:di}). As before, choose the
physical transverse polarisation vectors to be linear combinations of the
left and right circular polarisations, i.e. $a^\m = \a m^\m + \b \bar m^\m$. 
In this sector, the equation of motion becomes
\begin{equation}
\left(\matrix{k^2 - i \e_{\m\n\l\r} m^\m \bar m^\n k^\l L^\r &0 \cr
0 & k^2 -i \e_{\m\n\l\r} \bar m^\m m^\n k^\l L^\r \cr} \right)~
\left(\matrix{\a \cr \b \cr} \right) ~~=~~ \left(\matrix{0 \cr 0 \cr} \right)
\label{eq:dm}
\end{equation}
We see immediately, that in contrast to the `Weyl' case above, the symmetries
of the antisymmetric tensor ensure that this matrix is diagonal and therefore
the left and right circular polarisations are themselves the eigenstates, i.e.
the states which propagate with a well-defined velocity. We can readily check
that the light cone condition following from eq.(\ref{eq:dm}), viz.
\begin{equation}
k^2~~ =~~ \pm ~i \e_{\m\n\l\r}~ m^\m \bar m^\n k^\l L^\r
\label{eq:dn}
\end{equation}
reproduces eq.(\ref{eq:dk}).

As noted above, linearly polarised light may be considered as a superposition of the 
left and right circular polarisations with a phase difference which determines the 
direction of linear polarisation. If the left and right circular polarisations propagate 
at different speeds, then this phase difference will vary with the distance propagated,
thereby rotating the direction of linear polarisation. This birefringent rotation
angle is half the induced phase difference between the left and right circular
polarisations due to their velocity difference.
Linearly polarised light propagating through a distance $D$ will therefore experience
a birefringent rotation of the direction of polarisation given by
\begin{equation}
\d\phi ~~=~~  {2\pi D\over \l}~\d v
\label{eq:do}
\end{equation}
where $\d v$ is given by eq.(\ref{eq:dl}). 

We see, therefore, that there is a distinct difference between the birefringence effects
associated with the `Weyl' ($\Psi^{(K)}$) and Chern-Simons ($L^\m$) types of Lorentz
breaking. While the latter simply rotates the direction of linear polarisation, the
former in addition turns linear polarisation into elliptic polarisation.
Either effect is a clear signal for Lorentz violation; the presence of the latter is
characteristic of the parity and CPT violating Chern-Simons interaction. 

Clearly, since $\d v$ is expected to be tiny given that we know Lorentz invariance to be a 
good symmetry to high accuracy, in order to achieve measurable effects we need the longest 
possible baseline $D$, i.e.~we need to study deep astrophysical sources emitting linearly 
polarised radiation for which we can associate a definite direction. We now consider two
possible types of source.

\subsection{Radio galaxies}

A suitable set of astrophysical sources for testing this Lorentz-violation induced
birefringence are radio galaxies \cite{CFJ}. These emit synchrotron radiation, which is
linearly polarised. Moreover, for a large class of these objects, it is possible to 
identify an axis along which they are elongated, and models suggest that the alignment
of their magnetic field determines the orientation of the emitted linear polarisation
to be either along, or orthogonal to, this axis.

Carroll, Field and Jackiw\cite{CFJ} have carried out a detailed study of a sample of 
160 such radio galaxies, with redshifts typically in the range $z\sim 0.1$ to $z\sim 1$.
Making cuts to those with polarisations above 5\% and with $z > 0.4$, and after 
compensating for the effects of Faraday rotation by the intergalactic magnetised plasma,
they find clear evidence that the observed orientation of the linear
polarisation is indeed peaked around $90^o$ relative to the axis of the radio galaxy. 
The mean angle is found to be $90.4^o ~\pm~ 3.0^o$. They conclude that at the 95\%
confidence level, the birefringent rotation angle implied by a Chern-Simons mechanism
is $\d \phi < 6.0^o$.

To convert this into a bound for the Chern-Simons coupling, we need to convert from the
redshift to the proper distance $D$. For an $\Omega = 1$ matter dominated universe, this
gives $D = {2\over 3H_0} \Bigl(1 - (1+z)^{-{3\over2}}\Bigr)$, where $H_0$ is the Hubble
constant. Letting $h = H_0/100 ~{\rm km~sec}^{-1}~{\rm Mpc}^{-1}$, the recent WMAP
data\cite{WMAP} determines $h = 0.72 \pm 0.05$. 
It follows from eqs.(\ref{eq:dl}), (\ref{eq:do}) that
\begin{eqnarray}
\d \phi ~~=~~ {1\over2} ~D~L.\hat k ~~~~~~~~~~~~~~~~~~~~~~~\\
~~=~~ {1\over 3H_0}~\Bigl(1 - (1+z)^{-{3\over2}}\Bigr)~L.\hat k
\label{eq:dp} 
\end{eqnarray}
from which, taking $z\sim 0.4$, we deduce the following bound on the Chern-Simons 
coupling\cite{CFJ}
\begin{equation}
|L.\hat k| ~~<~~ 1.2 \times 10^{-42} ~{\rm GeV}
\label{eq:dq}
\end{equation}

Alternatively, from the discussion following eq.(\ref{eq:dh}), we can use the 
observation of linear (rather than elliptic) polarisation, with the predicted 
orientation, to estimate a bound for the dimensionless `Weyl' coupling $\Psi_0^{(K)}$ 
obtained from the couplings $K_{\m\n\l\r}$ in the phenomenological Lagrangian 
(\ref{eq:ca}). With the above bound on $\d\phi$, and a typical radio galaxy 
wavelength of $\l \sim 20$cm, we can use eq.(\ref{eq:do}) together with (\ref{eq:dh}) 
to obtain the following estimate:
\begin{equation}
|\Psi_0^{(K)}| ~~< ~~ O(10^{-29})
\label{eq:dr}
\end{equation}

Notice that because of the qualitatively different dispersion relations for the `Weyl'
and Chern-Simons mechanisms, i.e.~$\d v \sim {\rm const}$ for `Weyl' (eq.(\ref{eq:dh}))
and $\d v \sim {1/k}$ for Chern-Simons (eq.(\ref{eq:dl})), the birefringent rotation 
angle $\sim D \d v/\l$ is independent of the frequency of the radiation for Chern-Simons
but is proportional to frequency for the `Weyl' case, as used above.\footnote{It
should be noted, however, that the constant dispersion relation in the `Weyl' case
is a consequence of only retaining the renormalisable $dim = 4$ operator in the
extended QED Lagrangian. In the general case where higher-dimensional operators
are considered as well, as in the strong equivalence violating Lagrangian in section 2,
the corresponding dispersion relation is highly non-trivial\cite{Sfive}.}
It follows that in order to bound the `Weyl' coupling, we can look not only for sources
with long baselines $D$, but also study them at shorter wavelengths.

Taking advantage of this, a more recent estimate has been made by Kosteleck\'y 
and Mewes\cite{KM} using a smaller sample of higher redshift galaxies and quasars 
for which polarimetric measurements are available in the infra-red, optical and
UV. From this, they are able to quote an improved bound of $O(10^{-32})$ on a typical 
component of  $K_{\m\n\l\r}$.

\subsection{Gamma-ray bursts}

As noted above, one way to improve these bounds is to increase the baseline distance 
$D$ by studying polarisation in astrophysical sources at higher redshifts, while
for some couplings, observing sources at higher frequencies will also help. An 
intriguing new possibility in this direction has been opened up in the last few years 
by the detection of linear polarisation in gamma-ray burst (GRB) afterglows, i.e.~the 
fading optical phase observed after the initial prompt burst of gamma-rays, and
recently in the prompt gamma-ray phase itself.

The first observation of polarisation in GRB afterglows was made in 1999 in 
GRB990123, which has a redshift of $z=1.60$. Since then, around ten further observations
have been made. (See ref.\cite{CGLM} for a complete list.) Amongst these, the highest
redshift recorded so far is $z=2.33$, for GRB021004. 

It is very likely that in the next few years increasingly precise observations
of GRBs, notably with the forthcoming launch of the SWIFT satellite, will substantially
increase the data set of observations of polarisation in GRB afterglows. Eventually
we may hope that polarisation measurements on GRBs will evolve from their present 
role as probes of the basic mechanism of GRB production to tools which we can exploit
to constrain the fundamental physics of light propagation. If GRB polarisation
does indeed come to be as well understood as the polarisation from radio galaxies
discussed above, the increase in redshift by a factor of around 10 will allow us
to tighten the bounds (\ref{eq:dq}),(\ref{eq:dr}) still further. 

In order to realise this programme, we need first of all to observe intrinsic linear 
polarisation in high redshift GRBs, and also to understand its direction in terms
of the geometry of the source/observer so that we can place bounds on any birefringent
rotation. We also need to understand the effect on polarisation of the transmission
of the afterglow light through the host galaxy (see e.g. refs.\cite{LC,DDR}, 
a challenging task analogous 
to the subtraction of Faraday rotation in the intergalactic medium in the analysis 
of polarisation with radio galaxies. It will clearly be some time before all this
can be achieved, but our purpose in this section is simply to draw attention to the
possibility of using GRBs as controlled long-baseline light sources for the study
of fundamental physics such as the possible violation of Lorentz invariance.

As an illustration of how directed, linear polarisation can arise in GRB 
afterglows\footnote{In what follows, we are describing a mechanism for polarisation
in GRB afterglows, where it seems to be accepted that the relevant production 
mechanism is synchrotron radiation. Polarisation of the prompt gamma-rays in a GRB 
has recently been observed in GRB021206. Here, authors appear divided on whether the
origin is synchrotron radiation or inverse Compton scattering. See ref.\cite{DDR}
for a brief summary.}, we consider a simple model proposed by Ghisellini and 
Lazzati\cite{GL}. Several other models can be found in the literature, which differ
in important respects and should be readily distinguishable as more data becomes 
available.\footnote{We do not enter here into the comparison of models, which differ
in such important features as whether the magnetic field in the jet of ejected matter 
is completely tangled or partially ordered, whether the jet is homogeneous or
structured, and whether or not the jet cone opening angle is itself sufficiently
small that only the angle to the observer's line of sight is relevant. See, for
example, refs.\cite{Lazzati,DDR} for an introduction to the literature.} The key features
in many cases though are common:~
(i)~~GRBs arise through the ejection of a jet of relativistic matter
in a narrow cone from a central source, almost certainly a supernova. This jet 
structure removes spherical symmetry and provides the essential axis for the
source analogous to the elongation direction of the radio galaxies.~
(ii)~~Synchrotron radiation is emitted from the relativistic electrons
accelerated in the magnetic field of this jet, with the radiation emitted
at $90^o$ being linearly polarised.~
(iii)~~This synchrotron radiation is Lorentz boosted by the usual
relativistic aberration mechanism into the forward direction where it can be received
by the observer.~
(iv)~~Crucially, observations are made off the axis of the jet cone. 
This introduces the asymmetry necessary to observe a non-vanishing net polarisation. 
  
Consider then a shell of magnetised plasma moving outwards relativistically
in a jet making a cone of half-angle $\theta_c$. Synchrotron radiation is emitted
from the shell with the radiation emitted in its plane, i.e.~at $90^o$ to the outward
motion, being linearly polarised. This jet is observed off-axis by an observer
whose line of sight makes an angle $\theta_o$ with the jet axis. See Figure \ref{Fig:1}.
\FIGURE
{\epsfxsize=9cm\epsfbox{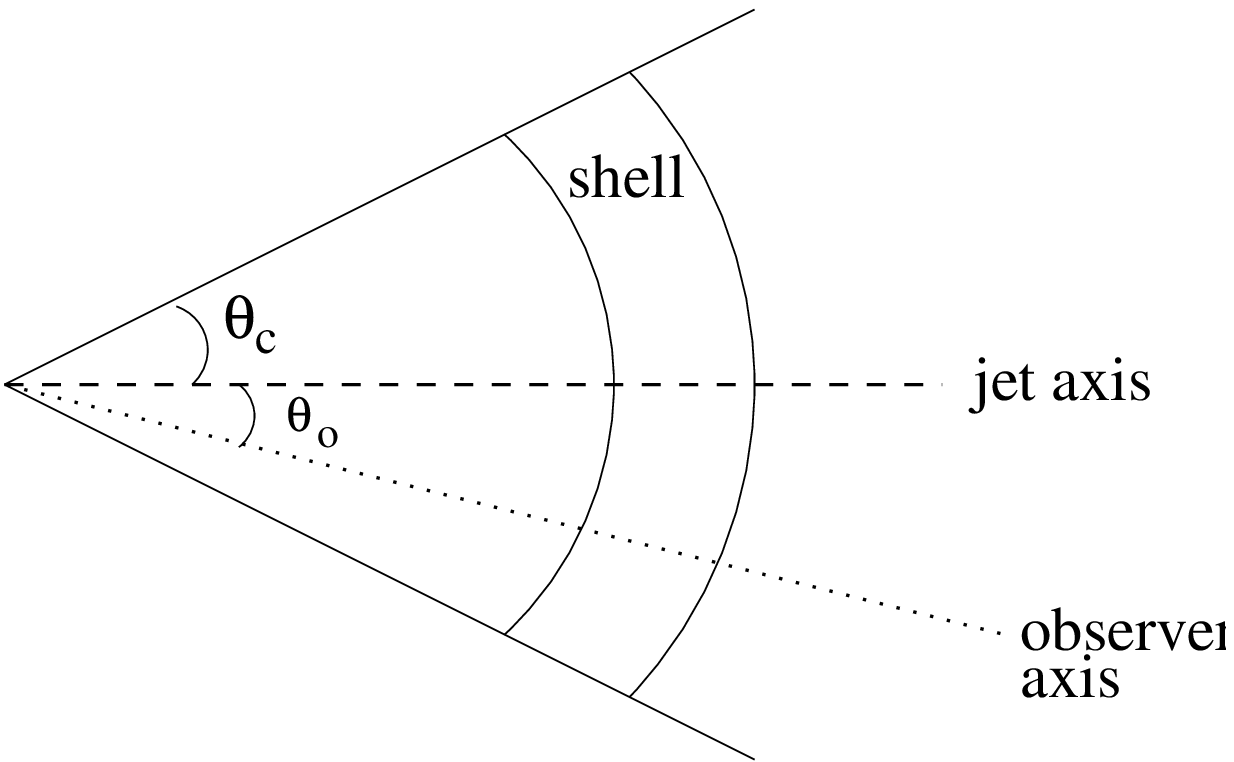}
\caption{A jet comprising a shell of magnetised plasma is emitted radially
outwards forming a cone of half-angle $\theta_c$. It is observed off-axis by an
observer whose line of sight makes an angle $\theta_0$ with the jet axis.}\label{Fig:1}}

Because of relativistic aberration, photons emitted in the plane of a segment of the shell,
in its comoving frame, are observed in the forward direction making an angle 
$\theta \sim 1/\c$ with the normal, where $\c$ is the Lorentz factor. See Figure \ref{Fig:2}.
\FIGURE
{\epsfxsize=9cm\epsfbox{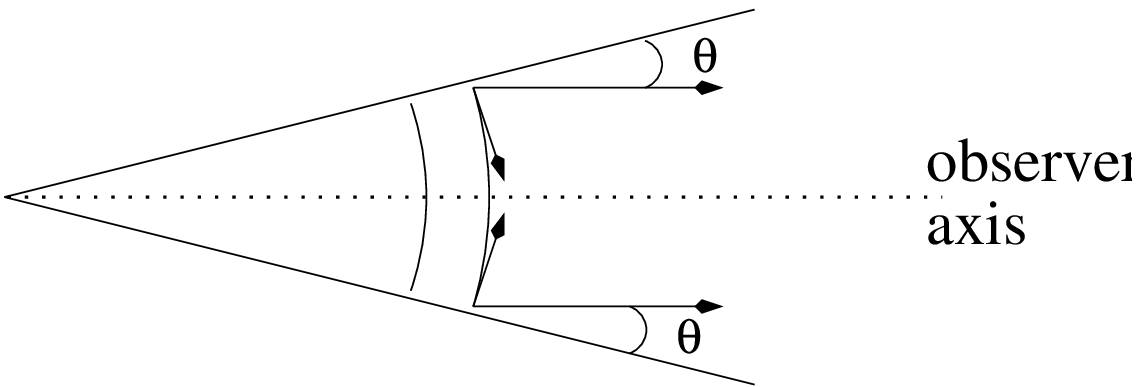}
\caption{Polarised light emitted at right angles to the outward radial motion
of a section of the plasma shell is relativistically aberrated into the forward direction, 
where it makes an angle $\theta \sim 1/\c$ with the normal to the shell. It is observed
at the edge of the circle subtending a cone of half-angle $\theta$ about the observer's
line of sight axis.}\label{Fig:2}}
This means that the observer receives linearly polarised light from the edge of the
circle subtending a cone of half-angle $\theta$ around the line of sight $\theta_o$.
The polarisation direction is radial with respect to this circle.

If the observer's cone lies entirely within the jet cone, then this whole circle is
observed and the net polarisation is zero through cancellation of the radially-directed
polarisation around the circle. As $\theta$ increases, however, only
a part of the observer's cone overlaps with the jet cone. In this case, only part of 
the circle is observed and a net polarisation is measured. See Figure 3. Finally, as $\theta$ 
increases still further, the observer's cone completely envelopes the jet cone and again no
polarisation is observed. 
\FIGURE
{\epsfxsize=5.5cm\epsfbox{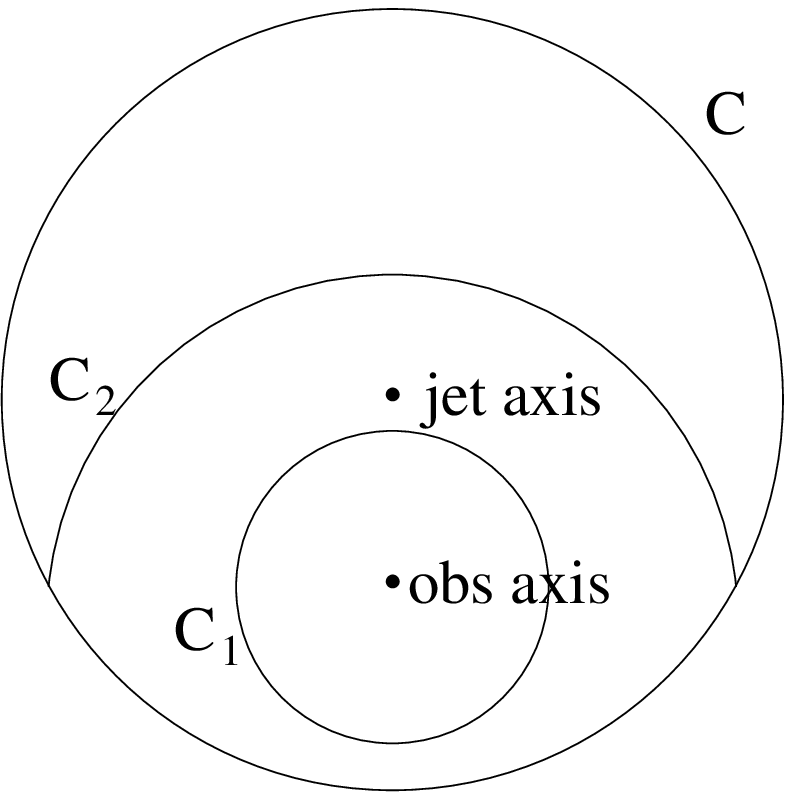}
\caption{The circle $C$ subtends the jet cone. Circles $C_1$ and $C_2$ subtend the `observer's
cone' of half-angle $\theta\sim 1/\c$ as $\theta$ increases. Initially $C_1$ lies entirely
within $C$, while for lower values of the Lorentz factor $\c$ there is only a partial
overlap, enabling a net linear polarisation to be observed.}\label{Fig:3}}

The criterion for observing a net linear polarisation is therefore
\begin{equation}
\theta_c + \theta_o ~~>~~ \theta ~~>~~ \theta_c - \theta_o
\label{eq:ds}
\end{equation}
that is,
\begin{equation}
{1\over \theta_c + \theta_o} ~~<~~ \c ~~<~~ {1\over \theta_c - \theta_o}
\label{eq:dt}
\end{equation}
Notice that this is only possible if $\theta_o \neq 0$, i.e.~if the GRB jet is
viewed off-axis. 

To summarise, at the beginning of the afterglow when $\c$ is large, the observer sees
only a fraction of the full jet ($\c > 1/\theta_c - \theta_o$) and no net polarisation
is observed. As $\c$ drops as the afterglow evolves, we enter the asymmetric stage
where the observer's cone overlaps partly with the jet cone -- this results in an
observed net polarisation. As $\c$ falls still further ($\c < 1/\theta_c + \theta_o$),
the entire jet cone becomes visible and the polarisation vanishes. 

This model also makes a definite prediction for the direction of polarisation. The 
derivation is given in detail in ref.\cite{GL} and consists of integrating the Stokes'
vectors Q and U around the visible arc of the observer's circle. The result, which
is a peculiarity of this precise geometric model, is that the polarisation is initially
orthogonal to the plane containing both the jet cone axis and the line of sight, 
before flipping abruptly to lie in this plane. All models of this type, however,
make a well-defined prediction of the direction of polarisation.

For our purposes, the crucial point to be extracted from this model is that we should
expect to observe a net linear polarisation, with a specified direction established with
respect to the plane formed by the jet axis and the observer's line of sight, in a
situation where a GRB jet is observed off-axis (as will generically be the case).
The existence, and possibly direction, of this linear polarisation is expected to
change with time as the Lorentz factor $\c$ reduces as the expanding jet slows. 

GRB afterglows therefore satisfy the criteria we need for the long-baseline 
astrophysical polarimetry measurements which can test Lorentz and CPT symmetry
to high accuracy. Moreover, if polarisation measurements meeting these criteria
can eventually also be made on the prompt gamma-rays themselves, we could also
improve the bounds on the `Weyl' coupling because of the extremely small wavelength
factor in the birefringent rotation formula, as discussed above. 

This combination of large redshifts and high frequencies means that GRBs have the 
potential ultimately to give the most accurate limits on the Lorentz and CPT violating 
couplings $|\Psi^{(K)}|$ and $|L^\m|$.

\section{Precision Atomic Spectroscopy}

While testing the Maxwell sector of the Lorentz and CPT violating theory requires 
observations on astronomical scales, testing the Dirac sector is the domain of
precision atomic physics experiments. In this section, we will consider the
implications of the phenomenological Lagrangian (\ref{eq:ca}) for precision
atomic spectroscopy, in particular with hydrogen and anti-hydrogen. The possibility
of making spectroscopic experiments on anti-hydrogen and looking for CPT violation
follows from the production for the first time of substantial numbers of cold 
anti-hydrogen atoms at the ATHENA\cite{ATHENA, ATHENAtwo} and ATRAP\cite{ATRAPone,ATRAPtwo} 
experiments at CERN in 2002. This subject has already been investigated in detail 
by Bluhm, Kosteleck\'y and Russell\cite{BKR}, but here we shall push 
their analysis a little further to see whether there are further options for CPT tests 
which could be realised at ATHENA and ATRAP. 

\subsection{Hydrogen spectrum with Lorentz and CPT violation}

Our starting point is the textbook derivation of the relativistic fine structure of
hydrogen energy levels starting from the Dirac equation, rather than the more
familiar derivation based on perturbation theory and the non-relativistic
Schrodinger equation. In outline, this derivation goes as follows. The non-relativistic
Schrodinger equation for hydrogen reduces to
\begin{equation}
\biggl[-{1\over 2m}\biggl({\partial^2\over\partial r^2} + 
{2\over r}{\partial\over\partial r} - {L^2\over r^2}
\biggr)~-~ {\a\over r} ~-~ E \biggr] ~\psi_{n\ell}({\underline r}) ~~=~~0
\label{eq:ea}
\end{equation}
where the wave function $\psi = \exp(-iEt) \psi_{n\ell}({\underline r})$,~ 
$L^2 \psi_{n\ell} = \ell(\ell + 1) \psi_{n\ell}$, and the energy eigenvalues
are $E = -{1\over2} m \a^2 /n^2$, where the integer $n$ satisfies 
$\ell = 0,1, \ldots, n-1$. 

We now reduce the full QED field theory to one-particle relativistic quantum mechanics
in the usual way, regarding the Dirac equation as a relativistic wave operator, 
in a classical background $A_\m$ field, acting on a four-component wave function
$\psi$, i.e.
\begin{equation}
\Bigl(i \c^\m D_\m ~-~m\Bigr)~\psi ~~=~~0
\label{eq:eb}
\end{equation}
where $D_\m = \partial_\m + i e A_\m$. Now act on this with the Dirac operator 
with $m \rightarrow -m$. This gives  
\begin{equation}
\Bigl( -D^2 - e \s^{\m\n} F_{\m\n}  - m^2 \Bigr)~\psi ~~=~~0
\label{eq:ec}
\end{equation}
Substituting the Coulomb potential $eA_0 = - {\a\over r}$,~$A_i = 0$, and using 
the explicit expressions for the $\c$ matrices\footnote{Our conventions for the 
$\c$ matrices are:
$$
\c^0 = \Bigl(\matrix{0&1\cr 1&0\cr}\Bigr) ~~~~~~~
\c^i = \Bigl(\matrix{0&\s^i\cr -\s^i&0\cr}\Bigr) ~~~~~~~
\c^5 = \Bigl(\matrix{-1&0\cr 0&1\cr}\Bigr)
$$
where $\underline \s$ are the Pauli matrices, and
$$
\s^{\m\n} = {i\over4} \bigl[\c^\m,\c^\n\bigr]
$$
which implies
$$
\s^{0i} = -{i\over2}\Bigl(\matrix{\s^i&0\cr 0&-\s^i\cr}\Bigr) ~~~~~~~
\s^{ij} = {1\over2}\e^{ijk}\Bigl(\matrix{\s_k&0\cr 0&\s_k\cr}\Bigr)
$$
We also need the following identity in the derivation of eq.(\ref{eq:ed}):
$$
\{\c^\l,\s^{\m\n}\} ~=~ - \e^{\l\m\n\r}\c^5\c_\r
$$
}, this becomes
\begin{equation}
\biggl[\biggl({\partial^2\over\partial r^2} + {2\over r}{\partial\over\partial r}\biggr)
- {L^2\over r^2} + \Bigl({\a\over r}\Bigr)^2 + 2E{\a\over r}  + E^2 - m^2
-i {\a\over r^2} \biggl(\matrix{\underline\s.\underline{\hat r} & 0 \cr 0 
&-\underline\s.\underline{\hat r}\cr}\biggr)~\biggr]~\psi ~~=~~0
\label{eq:ed}
\end{equation}
which separates into two almost identical equations for the two-component spinor 
wave functions in the decomposition $\psi = \Bigl(\matrix{\psi_+\cr \psi_-}\Bigr)$, viz.
\begin{equation}
\biggl[\biggl({\partial^2\over\partial r^2} + {2\over r}{\partial\over\partial r}\biggr)
-\Bigl\{L^2 - \a^2 \pm i\a \underline\s.\underline{\hat r}\Bigr\} {1\over r^2} 
+ 2E{\a\over r} + E^2 - m^2 \biggr]~\psi_{\pm} ~~=~~0
\label{eq:ee}
\end{equation}
We can show\cite{IZ} that the eigenvalues of the operator $\{L^2 - \ldots \}$
can be written as $\l(\l+1)$ where $\l = \bigl[(j+{1\over2})^2 - \a^2\bigr]^{1\over2},~~
\bigl[(j+{1\over2})^2 - \a^2\bigr]^{1\over2} - 1$ for $j = \ell \pm {1\over2}$ respectively.
In each case, this gives $\l = \ell-\d$ where $\d = {1\over 2j+1}~\a^2 + O(\a^4)$.

Now, comparing eq.(\ref{eq:ee}) with the non-relativistic Schrodinger equation 
(\ref{eq:ea}), we deduce by inspection that the energy eigenvalues in the 
relativistic equation satisfy
\begin{equation}
E^2 - m^2 ~~=~~ - {1\over (n-\d)^2}~ (E\a)^2
\label{eq:ef}
\end{equation}
Expanding in powers of $\a$, we recover the familiar expression for the fine-structure
correction to the hydrogen energy levels in the relativistic theory:
\begin{equation}
E_{\rm FS} ~~=~~ m ~-~ {1\over2}m\a^2 ~{1\over n^2} ~+~ 
{1\over2}m\a^4~{1\over n^3}~\Bigl({3\over4n} - {1\over j+{1\over2}}\Bigr) ~+~ O(\a^6)
\label{eq:eg}
\end{equation}

We now consider the implications of changing the Dirac equation to the extended one
implied by the Lorentz-violating Lagrangian (\ref{eq:ca}), viz.
\begin{equation}
\Bigl( i \c^\m D_\m - m - a_\m \c^\m - b_\m \c^5 \c^\m + i c_{\m\n} \c^\m D^\n + 
i d_{\m\n} \c^5 \c^\m D^\n - {1\over2} h_{\m\n} \s^{\m\n} \Bigr)~\psi ~~=~~ 0
\label{eq:eh}
\end{equation}
Multiplying as before by the equivalent operator with $m \rightarrow -m$, we find that the
extended Dirac one-particle relativistic wave equation analogous to eq.(\ref{eq:ec})
is
\begin{eqnarray}
\nonumber\\
&\biggl[ - D^2 - e\s^{\m\n}F_{\m\n} - m^2 - 2i a_\m D^\m - 4 b_\m \c^5 \s^{\m\l}D_\l
- c_{\m\n} \Bigl(2D^\m D^\n - ie F^{\m\n} - 2e \s^{\m\l}F_\l{}^\n \Bigr)~~~~~~ \nonumber\\
&- d_{\m\n}\c^5 \Bigl(-4i \s^{\m\l}D_\l D^\n - ie F^{\m\n} -2e \s^{\m\l}F_\l{}^\n \Bigr)
+ {i\over2} h_{\m\n} \e^{\m\n\l\r}\c^5 \c_\r D_\l \biggr]~\psi ~~=~~0 ~~~~~~~~~
\label{eq:ei}
\end{eqnarray}

To see the effect of the new Lorentz and CPT violating couplings, consider first just
adding the $a_\m$ and $b_\m$ terms. Using the explicit expressions for the $\c$ matrices,
we find, to first order in the small couplings $a_\m, b_\m$:
\begin{eqnarray}
\nonumber\\
&\biggl[\biggl({\partial^2\over\partial r^2} + {2\over r}{\partial\over\partial r}\biggr)
- {L^2\over r^2} + \Bigl({\a\over r}\Bigr)^2 + 2E{\a\over r}  + E^2 - m^2
- i {\a\over r^2} \biggl(\matrix{\underline\s.\underline{\hat r} & 0 \cr 0 &-
\underline\s.\underline{\hat r}\cr}\biggr) ~~~~~~~~~~~~~~~~~~~~~~~~~~~~ \nonumber\\
&+ 2E a_0 + 2a_0 {\a\over r}- {1\over2} E~ 
\underline b.\underline\s - {1\over2} {\a\over r}~ \underline b.\underline\s  
\biggr]~\psi  ~+~ 
O(\nabla\psi) 
~~=~~0
\label{eq:ej}
\end{eqnarray}
We now come to the key step. Comparing eqs.(\ref{eq:ed}) and (\ref{eq:ej}), we see
by inspection that to first order in the couplings, the energy eigenvalues are
simply related by
\begin{equation}
E~ \psi ~~\rightarrow~~ \Bigl(E ~+~ a_0 ~-~ {1\over4}~ \underline b. \underline \s~
\Bigr)~\psi
\label{eq:ek}
\end{equation}
That is, introducing the spin operator $\underline S = {1\over2} \underline \s$, the 
energy eigenvalues in the Lorentz violating theory are
\begin{equation}
E ~~=~~ E_{\rm FS} ~+~a_0 ~-~ {1\over2}~ \langle ~\underline b.\underline S~ \rangle
\label{eq:el}
\end{equation}
Extending the analysis to include the remaining couplings, we readily find the
complete result
\begin{equation}
E ~~=~~ E_{\rm FS} ~+~\bigl(a_0 + m c_{00}\bigr)~-~ 
{1\over2} \bigl( b_i - md_{i0} + \e_{ijk} h^{jk} \bigr)~\langle~ S^i~ \rangle 
\label{eq:em}
\end{equation}

To complete the derivation, we need the matrix element $\langle~ S^i~ \rangle$
taken in the basis of states $|n \ell s j m_j~\rangle$ appropriate for fine 
structure.\footnote{The evaluation of the matrix element of the spin operator
$\langle~ n \ell s j m_j|~ S_i~ |n \ell s j m_j~\rangle$
proceeds as follows. Abbreviate the notation $|n \ell s j m_j~\rangle$ to
$|j m_j~\rangle$. From the definition of the Clebsch-Gordon coefficients,
$$
|j m_j~\rangle ~~=~~ \sum_{m_\ell , m_s} ~|m_\ell m_s ~\rangle
~\langle~m_\ell m_s|j m_j~\rangle 
$$
we readily find
$$
\langle~ j' m'_j|~ S_z~ |j m_j~\rangle ~~=~~
\sum_{m_s} ~m_s~ \langle~j' m'_j|m_j-m_s,~ m_s ~\rangle~
\langle~m_j - m_s,~ m_s|j m_j~\rangle
$$
while
$$
\langle~ j' m'_j|~ S_{\pm}~ |j m_j~\rangle ~~=~~
\sum_{m_s} ~\sqrt{(s-m_s)(s+m_s+1)}~ \langle~j' m'_j|m_j-m_s,~ m_s + 1 ~\rangle~
\langle~m_j - m_s,~ m_s|j m_j~\rangle
$$
For $s={1\over2}$, and using the known expressions for the Clebsch-Gordon 
coefficients\cite{Merzbacher}, we then find for $m'_j = m_j$ and the two possibilities 
$j', j = \ell \pm {1\over2}$,
$$
\langle~ \ell \pm {1\over2}~ m_j|~ S_z~ |\ell \pm {1\over2}~ m_j~\rangle ~~=~~
\pm~{1\over 2\ell + 1} ~m_j ~~~~~~~~~~~~~~~~~ 
$$
$$
\langle~ \ell \pm {1\over2}~ m_j|~ S_z~ |\ell \mp {1\over2}~ m_j~\rangle ~~=~~
- {1\over 2\ell + 1}~\Bigl[\bigl(\ell + {1\over2}\bigr)^2 - m_j^2 \Bigr]^{1\over2}
$$
while $\langle ~S_{\pm}~\rangle ~=~0$. The diagonal matrix element is the result quoted
above.} In fact, this is well-known from the perturbative 
analysis of the Zeeman effect. The result is 
\begin{equation}
\langle~ n \ell s j m_j|~ S_z~ |n \ell s j m_j~\rangle ~~=~~ g~m_j
\label{eq:en}
\end{equation}
where $g$ is the Land\'e $g$-factor,
\begin{equation}
g ~~=~~ {1\over 2 j(j+1)}~\Bigl( j(j+1) + s(s+1) - \ell(\ell+1) \Bigr) 
\label{eq:eo}
\end{equation} 
while $\langle~ S_{\pm}~ \rangle = 0$.  For $s={1\over2}$,
\begin{equation}
\langle~ n \ell s j m_j|~ S_z~ |n \ell s j m_j~\rangle ~~=~~\pm~{1\over 2\ell +1}~m_j
\label{eq:ep}
\end{equation}
for the two possibilities $j = \ell \pm {1\over 2}$.
Putting all this together, we finally find the following result for the energy 
eigenvalues for hydrogen in the Lorentz and CPT violating phenomenological theory 
(\ref{eq:ca}):
\begin{equation}
E ~~=~~ E_{\rm FS} ~+~\bigl(a_0 + m c_{00}\bigr)~-~ 
{1\over2} \bigl( b_3 - m d_{30} + h_{12} \bigr)~\Bigl[ (\pm) {1\over 2\ell +1}~m_j \Bigr] 
\label{eq:eq}
\end{equation}
for $j= \ell \pm {1\over2}$.

The important new feature here compared with the result quoted in ref.\cite{BKR},
which was concerned purely with the $1s - 2s$ transition involving only $\ell = 0$
states, is the $\ell$ dependence in the general case arising from the Land\'e $g$-factor.

Notice also that since the contribution of the $a_\m$ and $c_{\m\n}$ couplings to the 
energy levels is independent of the angular momentum and spin quantum numbers, they do
not affect energy differences between states so play no role in determining the transition 
frequencies. The couplings which do have the potential to change the transition frequencies
are $b_\m, d_{\m\n}$ and $h_{\m\n}$. These are the parity-violating couplings which did
not have analogues in the strong equivalence violating QED Lagrangian discussed earlier
(see the dictionary in eq.(\ref{eq:cb}).) In order to have curvature-induced effects on
atomic spectra, we would need to go beyond the extended QED model of section 2, either
by introducing parity-violating couplings on a purely ad-hoc basis into the 
phenomenological Lagrangian or by embedding QED in an extended strong equivalence violating 
standard model with electroweak parity violation.

\subsection{$1s-2s$ and $2s-nd~~ (n\sim 10)$ transitions in H and anti-H}

In order to test for Lorentz and CPT violation in atomic spectroscopy, at least in 
the context of this model, we need to compare transition frequencies which are
sensitive to the couplings $b_3, d_{30}$ and $h_{12}$. One promising approach is to
compare the frequencies of the equivalent spectral lines in hydrogen and anti-hydrogen,
now that the abundant production of cold anti-hydrogen atoms at 
ATHENA\cite{ATHENA,ATHENAtwo} and ATRAP\cite{ATRAPone,ATRAPtwo} 
has made precision spectroscopy on anti-hydrogen feasible for the first time.

As noted in section 3, the couplings change in the following way under charge
conjugation: ~$a_\m \rightarrow - a_\m$,  $b_\m \rightarrow b_\m$,  
$c_{\m\n} \rightarrow c_{\m\n}$, $d_{\m\n} \rightarrow - d_{\m\n}$,
$h_{\m\n} \rightarrow - h_{\m\n}$. Transition frequencies sensitive to the combination
$\bigl(b_3 - m d_{30} + h_{12}\bigr)$ will therefore be different for hydrogen and
anti-hydrogen. Measuring such a difference in their spectra would be a clear signal
for Lorentz, and possibly CPT, invariance.

So far, attention has focused on the $1s-2s$ transition. This is a Doppler-free,
two-photon transition: two-photon since it violates the usual single-photon,
electric-dipole selection rule $\D \ell = 1$ and Doppler-free since the recoil
momentum from the emission of the two photons cancels. The $2s$ state is therefore
exceptionally long-lived (lifetime 122ms) and the $1s-2s$ spectral line has an 
ultra-narrow natural linewidth of 1.3Hz, giving a resolution of $O(10^{-15})$. 
($\l = 243$nm for the two-photon $1s-2s$ transition.) 
It has been suggested that an even more accurate measurement of the line centre
to $\sim 1$mHz, corresponding to an ultimate resolution of one part 
in $10^{18}$, is in principle attainable\cite{CEHHZ,Holzscheiter}. 
The $1s-2s$ transition is therefore favoured for precision spectroscopy.

The selection rules for two-photon transitions are derived in ref.\cite{Cagnac}.
The principle is straightforward. The transition amplitude depends on the product of 
two vector dipole operators, which is decomposed as the sum $\bigl(T_q^0 + T_q^2\bigr)$
of scalar and rank-two irreducible tensor operators, where the quantum numbers $q$
are correlated with the helicities of the emitted photons. For the special case of
$\ell=0$ to $\ell=0$ transitions, only the scalar operator $T_q^0$ can couple the two 
states and the resulting selection rule is 
\begin{equation}
\D j ~=~ 0, ~~~~~~~~~~\D m_j ~=~ 0 ~~~~~~~~~~~~(\ell=0~ \rightarrow ~\ell=0)
\label{eq:er}
\end{equation}

Applied to the $1s-2s$ transition, the important fact for us is the constraint
$\D m_j = 0$. According to eq.(\ref{eq:eq}), the $1s$ and $2s$ energy levels receive
the same corrections if they have the same $m_j$. The selection rule $\D m_j=0$
therefore ensures that the $1s-2s$ transition frequency is unchanged by the 
Lorentz and CPT violating couplings. 

To overcome this problem, Bluhm, Kosteleck\'y and Russell\cite{BKR} have made a detailed
study of the $1s-2s$ hyperfine Zeeman transitions for hydrogen and anti-hydrogen.
We shall only briefly summarise some of their results here; see ref.\cite{BKR} for
full details. In particular, they exploit the $n$-dependence of the hyperfine splitting
for a certain spin-mixed state to show that the corresponding $1s-2s$ hyperfine transition
receives a frequency shift due to the Lorentz and CPT violating couplings of
\begin{equation}
\d \n^{H} ~~\sim~~ \k~\Bigl( b_3^e - b_3^p - m_e d_{30}^e + m_p d_{30}^p 
- h_{12}^e + h_{12}^p \Bigr)
\label{eq:es}
\end{equation}
where $\k$ is a combination of magnetic field dependent mixing angles. Notice that we
have included the corresponding couplings for a modified Dirac equation for the proton
as well as the electron in eq.(\ref{eq:es}).
Since the corresponding hyperfine states in anti-hydrogen have the opposite spin 
assignments to hydrogen, the result in this case is
\begin{equation}
\d \n^{\bar H} ~~\sim~~ \k~\Bigl(-b_3^e + b_3^p - m_e d_{30}^e + m_p d_{30}^p 
- h_{12}^e + h_{12}^p \Bigr)
\label{eq:et}
\end{equation}
An observation of this hyperfine transition in hydrogen would therefore exhibit
sidereal variations because of the frame dependence implicit in the Lorentz violating 
couplings. Moreover, if we could compare with the equivalent transition in 
anti-hydrogen, there would be an instantaneous difference
\begin{equation}
\D \n^{\bar H} - \D \n^{H} ~~\sim~~ 2\k~\Bigl( b_3^e - b_3^p\Bigr)
\label{eq:eu}
\end{equation}
This would provide a direct measurement of the CPT-odd coupling $b_3$.
A positive observation would therefore be a signal of the violation of CPT symmetry. 
If $\D \n$ could be measured with an accuracy comparable to the ultimate line centre
of $1 {\rm mHz}$, this would place a bound of $|b_\m| < 10^{-27}$GeV on the
CPT-violating coupling\footnote{A compilation of similar bounds from a variety of
atomic physics experiments is given in ref.\cite{Bluhm}. See also ref.\cite{Mavro}
for a review invoking a broader range of Lorentz violating models.}. 

The magnetic field needed to resolve the hyperfine states is naturally present
because cold anti-hydrogen is produced in magnetic fields in the ATHENA and ATRAP
experiments and it is likely that spectroscopic measurements will be made on
anti-hydrogen atoms confined in a magnetic trap. (For details, see e.g.~ref.\cite{MMM}.)
However, the inhomogeneous nature of the trapping fields produces
Zeeman broadening of the spectral lines and limits the resolution that can actually
be achieved in practice. It is estimated in ref.\cite{BKR} that this could broaden
the $1s-2s$ hyperfine spectral line described above to over 1MHz width, which 
corresponds to an actual experimental resolution of only one part in $10^9$.
In ref.\cite{Holzscheiter}, the possibility of achieving a width of 20kHz is envisaged.
Reducing Zeeman broadening and developing techniques to determine the line centre
to even higher accuracy in order to approach the theoretically limiting resolution
of the $1s-2s$ transition is therefore an important experimental challenge\cite{MMM}.

In the remainder of this section, we will investigate an alternative to $1s-2s$ for
precision studies of Lorentz and CPT violation, exploiting the $\ell$-dependence in
the Land\'e $g$-factor in the formula (\ref{eq:eq}) for hydrogen energy levels.
Of particular interest is the Doppler-free, two-photon $2s-nd$ transition for $n\sim 10$.
This turns out to be surprisingly competitive with the $1s-2s$ transition for practical
precision spectroscopy. Indeed, at the time of writing of the comprehensive review
of hydrogen spectroscopy ref.\cite{Book}, the $2s-8,10,12d$ transitions provided the
most accurate determination of the Rydberg constant, to one part in $10^{10}$\cite{Lichten}. 
In fact, the limitation on further improvement comes not from the natural width of the
spectral lines but from the accuracy of the optical laser frequencies. The $2s-10d$
natural linewidth is 296kHz while the experimental linewidth reported in 
ref.\cite{Allegrini} is 1.25MHz. 

It seems possible, therefore, that in realistic experimental conditions, measurements
of the $2s-nd~~(n\sim10)$ transition in free hydrogen and anti-hydrogen (see 
ref.\cite{Haensch} for `atomic fountain' techniques to study free cold atoms outside
magnetic traps) could be made with a high precision complementing the $1s-2s$
hyperfine Zeeman transition\footnote{The case $n=11$ may be of particular interest
in the ATHENA programme as anti-hydrogen atoms in this state will be produced using
appropriate laser frequencies to induce transitions from the high$-n$ Rydberg states
in which they are originally formed. 
}. The loss of accuracy due to the broader natural linewidth
of $2s-nd$ is compensated by the absence of the Zeeman broadening which afflicts the
$1s-2s$ hyperfine transition, in addition to a variety of other experimental 
limitations including the accuracy of the 
laser frequencies at the required wavelengths. Naturally, though, a much more 
detailed experimental analysis would be required to confirm whether or not the
$2s-nd$ transition is really competitive, especially in the case of anti-hydrogen.

Our interest in the $2s-nd$ transition stems of course from the fact that invoking
an $\ell=2$ state allows us to exploit the $\ell$-dependence in the energy level formula
(\ref{eq:eq}). While the $2s$ state has only $j={1\over2},~m_j=\pm{1\over2}$ 
quantum numbers, the $nd$ state allows $j={3\over2},~m_j=\pm{3\over2},\pm{1\over2}$
and $j={5\over2},~m_j=\pm{5\over2},\pm{3\over2},\pm{1\over2}$.
Eq.(\ref{eq:eq}) therefore predicts a non-zero contribution from the Lorentz and
CPT violating couplings at the level of the basic spectral line, without needing
to introduce a magnetic field to produce hyperfine Zeeman splittings.

The selection rules for $2s-nd$ are as follows. For $\D \ell \neq 0$, the scalar 
operator $T_q^0$ gives no contribution, so the selection rules follow from the 
two-photon matrix elements of the rank-two tensor operator $T_q^2$. The selection 
rules for the $\D\ell=2$ transition are\cite{Cagnac}
\begin{equation}
|\D j| ~\le~2, ~~~~~~~~\D m_j ~=~q_1 + q_2  ~~~~~~~~~~(\D\ell=2)
\label{eq:ev}
\end{equation}
where $q_1,q_2$ are the photon helicities. Transitions between nearly all the 
$j,m_j$ states above are therefore possible, with the only restriction $|\D m_j|\le 2$. 

The frequency shift induced by the new couplings is therefore\footnote{For simplicity
of presentation, we are neglecting the effect of the nuclear proton spin here.}
\begin{equation}
\D\n ~~=~~ -{1\over2} \Bigl(b_3 - m d_{30} + h_{12}\Bigr)~
\Bigl[ \pm {1\over5} m_j^{\rm nd} ~-~ m_j^{2{\rm s}}\Bigr]
\label{eq:ew}
\end{equation}
where the $\pm$ refers to $j={5\over2}$ and ${3\over2}$ respectively. 
Exploiting this dependence on the quantum numbers, successive measurements on hydrogen
alone can therefore determine the combination $\bigl(b_3 - m d_{30} + h_{12}\bigr)$.
In addition, there will be sidereal variations due to the frame dependence associated
with Lorentz violation. 

The same measurements on anti-hydrogen would determine the charge conjugate contribution
$\bigl(b_3 + m d_{30} - h_{12}\bigr)$. Comparing the hydrogen and anti-hydrogen
results therefore enables us to make an independent determination of the CPT-violating
coupling $b_3$ and the Lorentz-violating, but CPT-preserving, combination 
$m d_{30} - h_{12}$. 

In conclusion, the transition $2s-nd$ in free hydrogen and anti-hydrogen allows us to
exploit in full the quantum number dependence of the general energy level formula
(\ref{eq:eq}). In realistic experimental conditions, taking account of issues such as 
Zeeman broadening in the magnetic trap required to study the hyperfine Zeeman $1s-2s$ 
transition and the limitations from laser frequency accuracy at the relevant wavelengths, 
it may even prove comparable in resolution with the $1s-2s$ transition. In any case,
it provides an interesting complementary measurement in anti-hydrogen spectroscopy.

\section{Conclusions}

In this paper, we have investigated the construction and experimental implications
of phenomenological extensions of QED exhibiting violations of the strong equivalence
principle, Lorentz invariance or CPT symmetry. The underlying theme has been the formal 
similarity between these models. Both are essentially theories of the phenomenology
of Lorentz non-invariant operators. In Kosteleck\'y's Lorentz-violating model, these 
operators arise with multi-index couplings (perhaps to be interpreted as VEVs of tensor
operators in a more fundamental theory of spontaneous Lorentz symmetry breaking), which 
may be bounded by experiment. In the strong equivalence violating models, this role
is played by the curvature tensors of the background gravitational field.

Modifications to the Maxwell sector of QED can be tested to high accuracy using 
long-baseline polarimetry observations of astrophysical sources. After reviewing existing
limits on birefringent propagation obtained from radio galaxies and quasars, we have shown 
how in future polarisation measurements on gamma-ray bursts at high redshift
have the potential to enable even more stringent bounds to be placed on the 
Lorentz-violating Maxwell couplings $K_{\m\n\l\r}$ and $L^\m$.

In the Dirac sector, experimental tests of Lorentz and CPT violation can be made
with high precision atomic spectroscopy comparing hydrogen and anti-hydrogen,
the latter having become feasible following the recent production of cold
anti-hydrogen atoms in significant numbers by the ATHENA and ATRAP collaborations.
On the basis of an extended formula for atomic energy levels, we have proposed
that the transition $2s-nd ~(n\sim 10)$ in free hydrogen and anti-hydrogen may provide
a useful complement to the currently envisaged programme of spectroscopy on the
$1s-2s$ hyperfine Zeeman transition.

One of the original motivations for this paper was to investigate whether the 
constraints on these models obtained from astrophysics could be used to bound the 
couplings giving rise to Lorentz and CPT violation in atomic spectroscopy, and 
vice-versa. At first sight, it appears that, in particular, the Chern-Simons
photon coupling $L^\m$ and the CPT-violating electron coupling $b_\m$ should be 
related through radiative corrections. If so, this would have important implications
for the atomic physics experiments since, as we have seen, the existing bounds 
from radio galaxies constrain $|L^\m| < O(10^{-42})$GeV while, even under the most
optimistic experimental scenario, anti-hydrogen spectroscopy can only fix a
bound of $|b_\m| < O(10^{-27})$GeV. However, ambiguities in the specification of the
quantum theory based on the Kosteleck\'y Lagrangian arising from the axial anomaly 
make the relation between $b_\m$ and $L^\m$ strictly indeterminate, though perhaps 
it could still be argued that we would naturally expect them to be of the same 
order of magnitude. On the other hand, where the relation between couplings is
free of anomaly ambiguities, as we would expect to be the case with $a_\m, c_{\m\n}$ 
and $K_{\m\n\l\r}$, the Dirac couplings do not affect the frequency of atomic
transitions. 

The strong equivalence principle, Lorentz invariance and CPT symmetry are all
fundamental ingredients of conventional quantum field theories of elementary particle
physics and cosmology. Nevertheless, as we have seen, it is relatively straightforward
to modify the dynamics of QED or the standard model to exhibit violations of each of
these principles and to predict corresponding experimental signatures, either in 
atomic physics or astronomy. Hopefully, these theoretical considerations will help 
to stimulate further high-precision experimental work to test their validity to still 
greater accuracy.

\acknowledgments

I would like to thank M. Charlton and D.P. van der Werf for interesting discussions.

\vfill\eject


\begin{thebibliography}{999}

\bibitem{DH}{I.T. Drummond and S. Hathrell, Phys. Rev. D22 (1980) 343.}
\bibitem{Sone}{R.D. Daniels and G.M. Shore, Nucl. Phys. B425 (1994) 634. }
\bibitem{Stwo}{R.D. Daniels and G.M. Shore, Phys. Lett. B367 (1996) 75.}
\bibitem{Sthree}{G.M. Shore, Nucl. Phys. B460 (1996) 379.}
\bibitem{Sfour}{G.M. Shore, Nucl. Phys. B605 (2001) 455.}
\bibitem{Sfive}{G.M. Shore, Nucl. Phys. B633 (2002) 271.}
\bibitem{Sseven}{G.M. Shore, {\it Superluminal Light}, to appear in the Proceedings:
`Time and Matter: An International Colloquium on the Science of Time', Venice 2002,
eds. I. Bigi and M. Faessler, gr-qc/0302116.} 
\bibitem{Ssix}{G.M. Shore, Nucl. Phys. B646 (2002) 281.}
\bibitem{Leon}{M.A. Leontovich, {\it in} L.I. Mandelshtam,
{\it Lectures in Optics, Relativity and Quantum Mechanics}, Nauka, Moscow 1972~
{(\it in Russian).}}
\bibitem{Ohkuwa}{Y. Ohkuwa, Prog. Theor. Phys. 65 (1981) 1058.}
\bibitem{BG}{F.A. Berends and R. Gastmans, Annals of Physics 98 (1976) 225.}
\bibitem{Milton}{K.A. Milton, Phys. Rev. D15 (1977) 538.}
\bibitem{DHGK}{J.F. Donaghue, B.R. Holstein, B. Garbrecht and T. Konstandin,
Phys. Lett. B529 (2002) 132.}
\bibitem{CK}{D. Colladay and V.A. Kosteleck\'y, Phys. Rev. D55 (1997) 6760; 
D58 (1998) 116002. }
\bibitem{Kreview}{V.A. Kosteleck\'y, in Proc. {\it Intnl. Conf. Orbis Scientiae 2000}, 
Coral Gables, Dec. 2000.}
\bibitem{Weinberg}{S. Weinberg, {\it The Quantum Theory of Fields}, Cambridge 
University Press, 1996.}
\bibitem{CFJ}{S.M. Carroll, G.B. Field and R. Jackiw, Phys. Rev. D41 (1990) 1231.}
\bibitem{KM}{V.A. Kosteleck\'y and M. Mewes, Phys. Rev. Lett. 87 (2001) 251304.}
\bibitem{JK}{R. Jackiw and V.A. Kosteleck\'y, Phys. Rev. Lett. 82 (1999) 3572.}  
\bibitem{ATHENA}{M. Amoretti {\it et al.}, Nature 419 (2002) 456.}
\bibitem{ATHENAtwo}{M. Amoretti {\it et al.}, Phys. Lett. B578 (2004) 23.}
\bibitem{ATRAPone}{G. Gabrielse {\it et al.}, Phys. Rev. Lett. 89 (2002) 213401.}
\bibitem{ATRAPtwo}{G. Gabrielse {\it et al.}, Phys. Rev. Lett. 89 (2002) 233401. }
\bibitem{BKR}{R. Bluhm, V.A. Kosteleck\'y and N. Russell, Phys. Rev. Lett. 79 (1997) 1432.}
\bibitem{BGVZone}{A.O. Barvinsky, Yu.V. Gusev, G.A. Vilkovisky and V.V. Zhytnikov,
Print-93-0274 (Manitoba), 1993.}
\bibitem{BGVZtwo}{A.O. Barvinsky, Yu.V. Gusev, G.A. Vilkovisky and V.V. Zhytnikov,
J. Math. Phys. 35 (1994) 3525; J. Math. Phys. 35 (1994) 3543; 
Nucl. Phys. B439 (1995) 561.}
\bibitem{Seight}{G.M. Shore, Contemp. Phys. 44 (2003) 503.} 
\bibitem{JW}{F.A. Jenkins and H.E. White, {\it Fundamentals of Optics}, McGraw-Hill, 
N.Y. 1976.}
\bibitem{WMAP}{D.N. Spergel {\it et al.}, WMAP Collaboration, Astrophys. J. Suppl. 148
(2003) 175.}
\bibitem{CGLM}{S. Covino, G. Ghisellini, D. Lazzati and D. Malesani, {\it Polarization
of Ggmma-ray burst optical and near-infrared afterglows}, astro-ph/0301608.}
\bibitem{LC}{D. Lazzati {\it et al.}, {\it Intrinsic and dust-induced polarization in
gamma-ray burst afterglows: the case of GRB021004}, astro-ph/0308540.}
\bibitem{DDR}{S. Dado, A. Dar and A. De Rujula, {On the polarization of gamma ray
bursts and their optical afterglows}, astro-ph/0403015.}
\bibitem{GL}{G. Ghisellini and D. Lazzati, {\it Polarization lightcurves and position
angle variation of beamed gamma-ray bursts}, astro-ph/9906471.}
\bibitem{Lazzati}{D. Lazzati, {\it Linear Polarization on Gamma-Ray Bursts: from the
prompt to the late afterglow}, astro-ph/0312331.}
\bibitem{IZ}{C. Itzykson and J-B. Zuber, {\it Quantum Field Theory}, McGraw-Hill, N.Y. 1985.}
\bibitem{Merzbacher}{E. Merzbacher, {\it Quantum Mechanics}, Wiley, N.Y. 1970.}
\bibitem{CEHHZ}{M. Charlton, J. Eades, D. Horvath, R.J. Hughes and C. Zimmermann,
Phys. Rep. 241 (1994) 65.}
\bibitem{Holzscheiter}{M.H. Holzscheiter {\it et al}, Nucl. Phys. B (Proc. Suppl.) 56A
(1997) 336.}
\bibitem{Cagnac}{B. Cagnac, G. Grynberg and F. Biraben, {\it Spectroscopie D'Absorption 
Multiphonique Sans Effet Doppler}, Journal de Physique 34 (1973) 845. (In French)}
\bibitem{Bluhm}{R. Bluhm, {\it Probing the Planck scale in low-energy atomic physics},
hep-ph/0111323.}
\bibitem{Mavro}{N.E. Mavromatos, {\it Theoretical and Phenomenological Aspects 
of CPT Violation}, hep-ph/0305215.}
\bibitem{MMM}{M.H. Holzscheiter, M. Charlton and M.M. Nieto, Phys. Rep. ({\it in press}).}
\bibitem{Book}{G.F. Bassani, M. Inguscio and T.W. H\"ansch, eds., {\it The Hydrogen
Atom}, Proceedings, Pisa Symposium, Springer-Verlag, Berlin, Heidelberg, New York, 1988.}
\bibitem{Lichten}{W. Lichten, {\it Hydrogen Spectroscopy and Fundamental Physics},
in ref.\cite{Book} 39-48.}
\bibitem{Allegrini}{M. Allegrini {\it et al.}, {\it Doppler-Free Two-Photon Spectroscopy
of Hydrogen Rydberg States: Remeasurement of $R_{\infty}$}, in ref.\cite{Book} 49-60.}
\bibitem{Haensch}{T.W. H\"ansch, {\it High Resolution Spectroscopy of Hydrogen}, in 
ref.\cite{Book} 93-102.}




\end{thebibliography}
\end{document}